\documentclass[sigconf,nonacm,printcopyright=False]{acmart}

\usepackage{booktabs} 
\usepackage{amsmath}
\usepackage{algorithm,algpseudocode}
\usepackage{amsmath}
\usepackage{amsfonts}
\usepackage{amssymb}
\usepackage{bbding}
\usepackage{pifont}
\usepackage{wasysym}
\usepackage{amssymb}
\usepackage{balance}

\renewcommand\footnotetextcopyrightpermission[1]{} 
\begin{document}

\setlength{\abovedisplayskip}{0pt}
\setlength{\belowdisplayskip}{2pt}

\title{NeuroMAX: A High Throughput, Multi-Threaded, Log-Based Accelerator for Convolutional Neural Networks
}

\author{Mahmood Azhar Qureshi}
\affiliation{%
 \institution{Kansas State University}
 \city{Manhattan}
 \state{Kansas}
}
\email{mahmood102@ksu.edu}

\author{Arslan Munir}
\affiliation{%
 \institution{Kansas State University}
\city{Manhattan}
 \state{Kansas}
}
\email{amunir@ksu.edu}

\begin{abstract}
Convolutional neural networks (CNNs) require high throughput hardware accelerators for real time applications owing to their huge computational cost. Most traditional CNN accelerators rely on single core, linear processing elements (PEs) in conjunction with 1D dataflows for accelerating convolution operations. This limits the maximum achievable ratio of peak throughput per PE count to unity. Most of the past works optimize their dataflows to attain close to a $100\%$ hardware utilization to reach this ratio. In this paper, we introduce a high throughput, multi-threaded, log-based PE core. The designed core provides a $200\%$ increase in peak throughput per PE count while only incurring a $6\%$ increase in area overhead compared to a single, linear multiplier PE core with same output bit precision. We also present a 2D weight broadcast dataflow which exploits the multi-threaded nature of the PE cores to achieve a high hardware utilization per layer for various CNNs. The entire architecture, which we refer to as NeuroMAX, is implemented on Xilinx Zynq 7020 SoC at 200 MHz processing clock. Detailed analysis is performed on throughput, hardware utilization, area and power breakdown, and latency to show performance improvement compared to previous FPGA and ASIC designs.
\vspace{-0.5 mm}
\end{abstract}

\vspace{-3 mm}

\maketitle

\vspace{-3 mm}
\section{Introduction}
\vspace{-1 mm}
Convolutional neural networks (CNNs) enable embedding of AI into devices for vision-based applications with an unprecedented accuracy. The early proposed high accuracy CNNs \cite{alexnet,vgg16,googlenet} required tens of millions of parameters and computations for one inference pass. This computational complexity along with high memory requirements greatly hampered their deployment on low energy, resource constrained devices. In addition to this, many CNN architectures used varying kernel sizes which results in reconfigurability requirement as well as low hardware utilization in accelerator designs. Separable convolution for CNNs was introduced the first time in mobilenets \cite{mobnetv1,mobnetv2} to reduce the number of multiply and accumulates (MACs). In addition, many modern CNNs use kernels of size $3\times3$ to promote ease of accelerator design with high hardware utilization and throughput. \par

Design of an efficient dataflow for scheduling data into the accelerator is equally important. An inefficient dataflow results in reduced hardware utilization which causes a decrease in throughput. Dataflow should also promote the reusability of data since, in most cases, the same kernels are being applied on the entire input feature map.
It has been shown previously that the movement of data to/from DDR memory is 200$\times$ more costly in terms of energy consumption than a standard MAC operation \cite{energy}. Thus, the dataflow design should not only optimize the throughput and area, but also the data movement, in order to ensure reduced energy expenditure.\par

Log-based accelerators have recently gained quite a lot of traction because of their simpler structure as compared to traditional accelerators with linear processing elements (PEs). Each PE in traditional CNN accelerator cores is essentially responsible for one multiplication in convolution operation. Log PEs replace the bulky multiplier cores with low \textit{cost} barrel shifters without incurring a significant loss in accuracy. We clarify that cost here primarily refers to the area cost, which is determined by the number of LookUp Tables (LUTs) for field-programmable gate arrays (FPGAs) and gate count for application-specific integrated circuits (ASICs). This area cost is important because there are limited resources on-chip and thus this area cost also translates to monetary cost of system-on-chip (SoC).
Many past approaches have designed log-based PE elements but have not exploited the low cost and overhead of such PEs. They instead rely on already established spatial architectures and 1D dataflows used for linear PEs. Our proposed \textbf{NeuroMAX} accelerator core comprises of 108 PEs arranged in a $6\times3\times6$, 3D spatial grid. The presented accelerator optimizes the most commonly used $3\times3$ and $1\times1$ kernel sizes to achieve high throughput and utilization. It can also be used for larger kernel sizes because of its grid structure and configurable 2D dataflow. Our main contributions are as follows:  
\vspace{-1.5 mm}
\begin{itemize}
\item We design a multi-threaded, low cost, log-based PE core. Using this core, we generate a spatial grid of 108 PEs, capable of performing a wide variety of convolution operations with high hardware utilization. \vspace{-0.5 mm}
\item We develop a 2D dataflow which exploits the thread based PE design to maximize the throughput and enhance data reuse to minimize the DDR memory access. \vspace{-0.5 mm}
\item We implement the entire NeuroMAX architecture in an FPGA and show improved performance in terms of area, throughput, hardware utilization, latency and power efficiency compared to past approaches.  
\end{itemize}
\vspace{-2 mm}
\begin{figure*}
\centering
\includegraphics[width=0.8\textwidth,keepaspectratio]{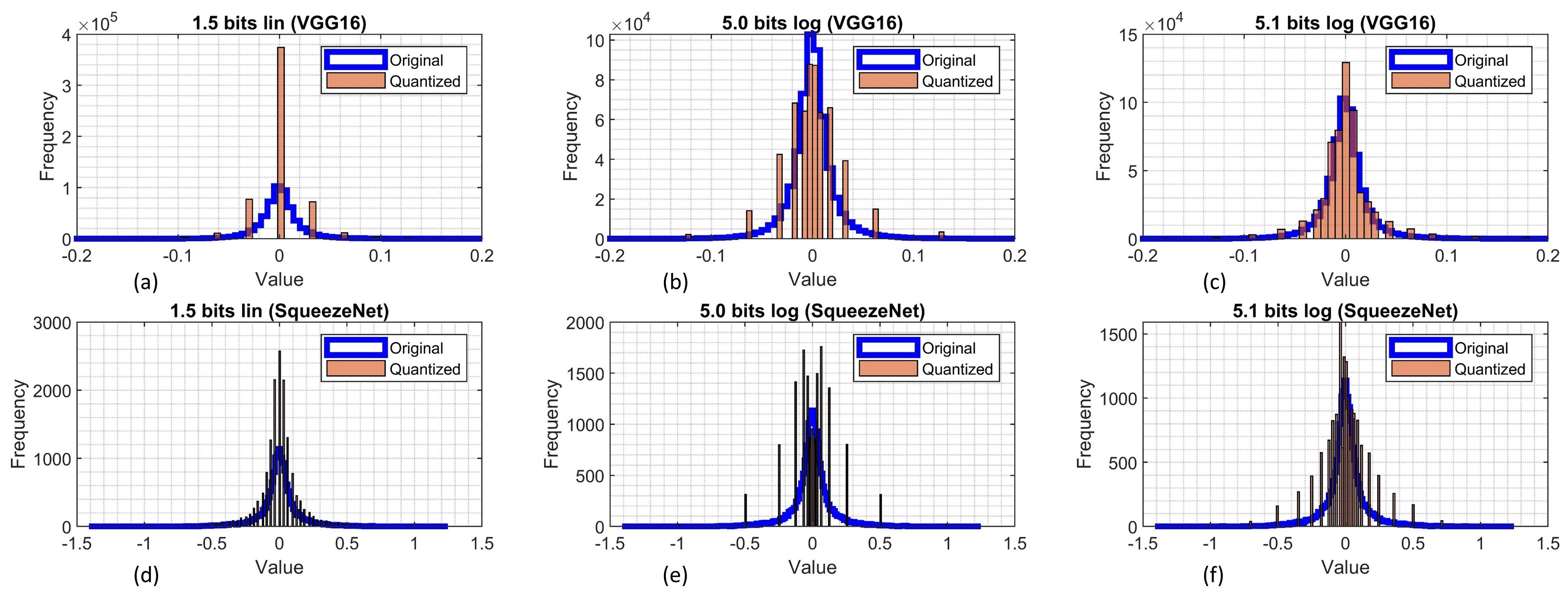}
\vspace{-3 mm}
\caption{Linear vs. Log Quantization (a) 1.5 bits linear VGG16 net (b) 5.0 bits log VGG16 (c) 5.1 bits log VGG16 (d) 1.5 bits linear SqueezeNet (e) 5.0 bits log SqueezeNet (f) 5.1 bits log SqueezeNet   }
\label{QuantFig}
\vspace{-5 mm}
\end{figure*}
%

\vspace{-1 mm}
\section{Related Work}
\vspace{-1 mm}
Many hardware accelerators have been proposed recently and in the past. \cite{eyeriss} proposed a non-systolic array, reconfigurable spatial architecture along with a new dataflow scheme called \textit{row stationary} to maximize the data reuse. However, this design incurs high PE cost owing to local storage and control in PE. It also has low hardware utilization which results in low throughput per PE. \cite{fpgaCNN1} proposes an FPGA-based CNN accelerator with integrated depth-wise separable mode of operation. This accelerator, however, has low throughput because of the usage of 32-bit floating point format. \cite{fpgaCNN2} proposes an FPGA-based CNN accelerator having a dedicated matrix multiplication engine (MME) on Arria 10 SoC. It achieves a frame rate of 266 fps, however, its MME engine has a huge digital signal processing (DSP) cost of 1200+ DSP blocks. \cite{eyerissv2} is the improved version of \cite{eyeriss} with higher hardware utilization and throughput. \cite{logquant1} introduced the concept of logarithmic data representation for neural network accelerator designs. It also gives accuracy comparison between linear and log quantization.
\cite{logquant2} proposes an accelerator design using arbitrary log base. It, however, does not utilize the low hardware overhead of the log-based PE and instead rely on linear PE arrangements. \cite{3dTiled} proposes a reconfigurable design for various convolution kernels. It uses a propagated input data flow scheme but incurs high latency and low hardware utilization. \cite{cnn5} proposes a rescheduled dataflow for convolution to optimize the energy efficiency. \cite{VWA} proposes a vectorwise accelerator architecture with the goal of maximizing the hardware utilization. It supports various kernel sizes from $1\times1$ to $5\times5$. \par 
Although some of the recent designs achieve  high hardware utilization, they are not able to increase the peak throughput per PE count beyond unity owing to the use of single core, linear PEs with high area cost.
This paper overcomes the limitations of prior works by leveraging \textit{log} PEs with multiple low cost threads within each log PE, and designing a 2D dataflow which promises high throughput by exploiting multi-level parallelism.

\vspace{-4 mm}
\section{Log Mapping}
\vspace{-1 mm}
Log mapping or log quantization maps an input value $x$ to a logarithmically quantized value $x'$. Many trained neural nets have weights \textit{w} and input activations \textit{a} which are non-uniformly distributed. Mapping these 32-bit floating point (fp32), non-uniformly distributed values over fixed point, linearly quantized values introduces significant amount of quantization noise for small bit width. Most hardware platforms use fixed point arithmetic for data manipulation where the fixed point number is represented in signed $Qm.n$ format. Here, $m$ represents the integer part whereas $n$ represents the fractional part. The range of values which can be represented are $range_{lin} = [-2^{m-1},2^{m-1}-\epsilon]$ where, $\epsilon = 2^{-n}$, is the step size.
\par
A linear quantizer rounds the fp32 value to the nearest multiple of $\epsilon$ and then clips it as follows:
%
\begin{equation}
\vspace{-0.5 mm}
x_q = clip\left[\left(round(\frac{x}{\epsilon})\right)\cdot \epsilon , -2^{m-1},2^{m-1}-\epsilon \right] 
\end{equation}

where,
\vspace{-1 mm}
\begin{equation}
\vspace{-1 mm}
     clip(x, min,max) = \begin{cases} \mbox{max,} & \mbox{} x \geq max \\ \mbox{$x$,} & \mbox{} min < x < max \\\mbox{min,} & \mbox{otherwise} \end{cases} 
\end{equation}

A log quantizer takes as input, $x$ and the quantization parameters $< m, n, b >$, where $b$ is the logarithmic base, and produces a log quantized value $x'$ as output. The quantization process can be written as:
\begin{equation}
\vspace{-0.5 mm}
x' = clip\left[\left(round({log_b(|x|)})\right), -2^{m-1},2^{m-1}-\epsilon \right] 
\vspace{-1 mm}
\end{equation}

\begin{equation}
\vspace{-0.5 mm}
     x_q = 
     \begin{cases} 
     0, & x = 0 \\ 
     sign(x)\cdot b^{x'}, & otherwise 
     \end{cases} 
     \label{quantize}
\end{equation}

Figure \ref{QuantFig} shows some of the quantization results for the first five convolution layers of VGG16 \cite{vgg16} and SqueezeNet \cite{squeezenet}. Instead of using log base-2 for quantization, we use log base-$\sqrt{2}$ for more accurate mapping as shown in Figure \ref{QuantFig}(c) and (f). Infact, we observe that VGG16, pretrained on ImageNet dataset, with fp32 data, after base-$\sqrt{2}$ quantization, has top-1 accuracy decrement by only $\approx$3.5\% from $67.5\%$ to $63.8\%$. This is opposed to log base-2 quantization which decreases the accuracy by $\approx$10\%. This observation has also been verified in \cite{logquant1}.

\begin{figure}
\vspace{-2 mm}
\includegraphics[width=0.35\textwidth,keepaspectratio]{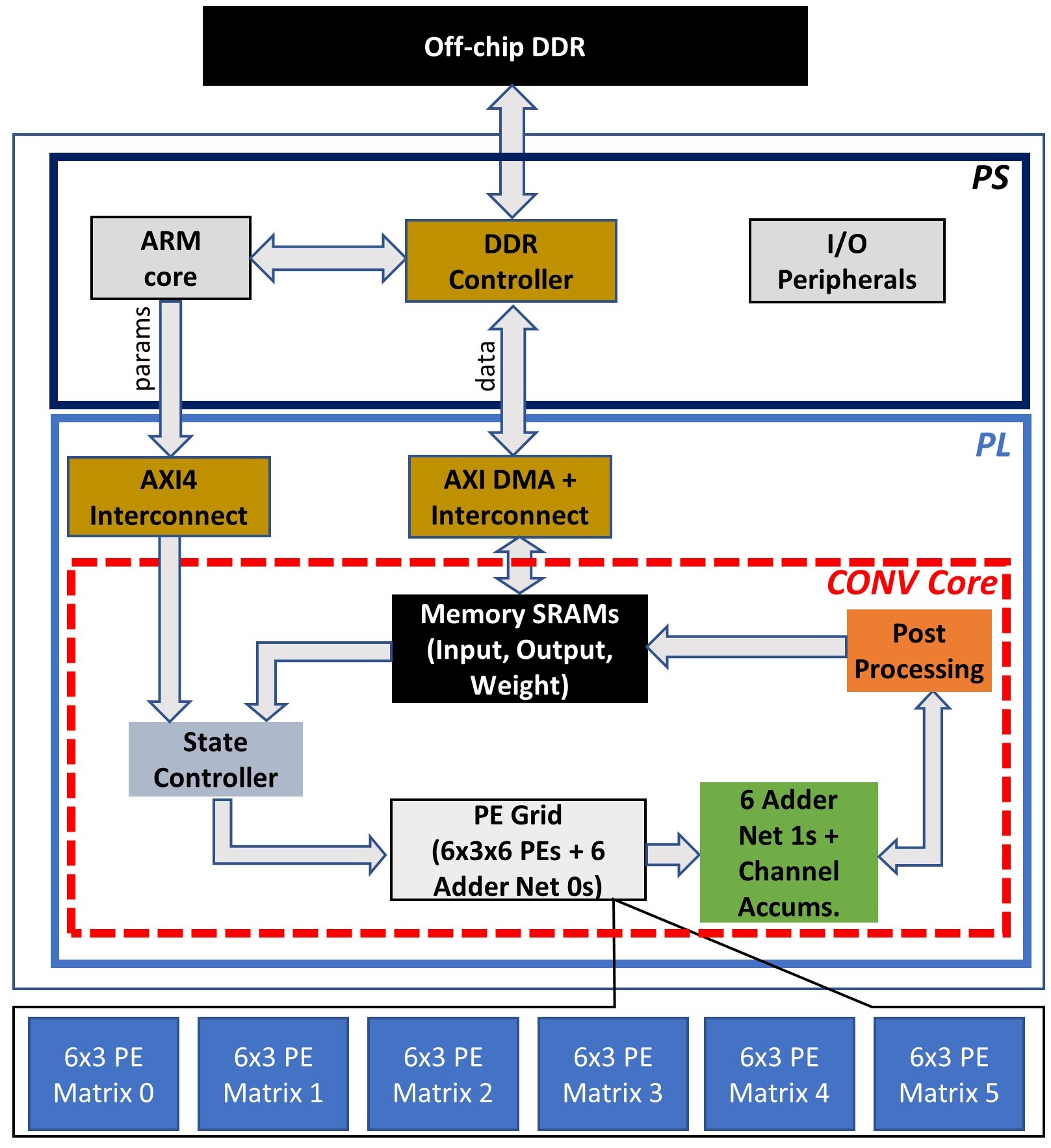}
\vspace{-3 mm}
\caption{NeuroMAX System Architecture}
\label{arch}
\vspace{-6 mm}
\end{figure}

\begin{figure*}
\centering
\includegraphics[width=0.80\textwidth,keepaspectratio]{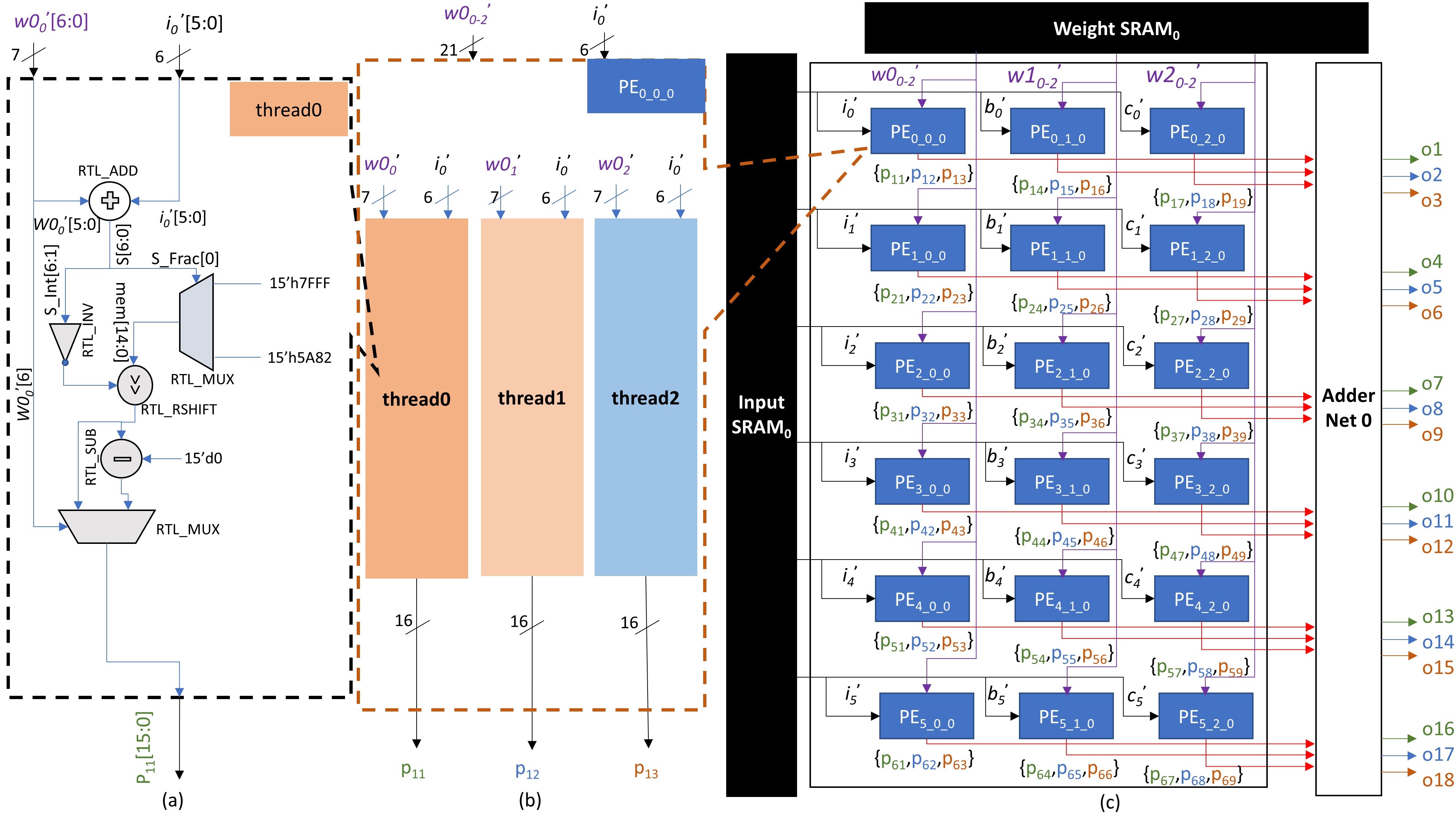}
\vspace{-4 mm}
\caption{(a) Compute Thread (b) Collection of Threads to Make a PE (d) 6x3 PE Matrix 0 and Adder Net 0}
\label{PEmatrix}
\vspace{-3 mm}
\end{figure*}

\begin{figure}
\vspace{1 mm}
\includegraphics[width=0.45\textwidth,keepaspectratio]{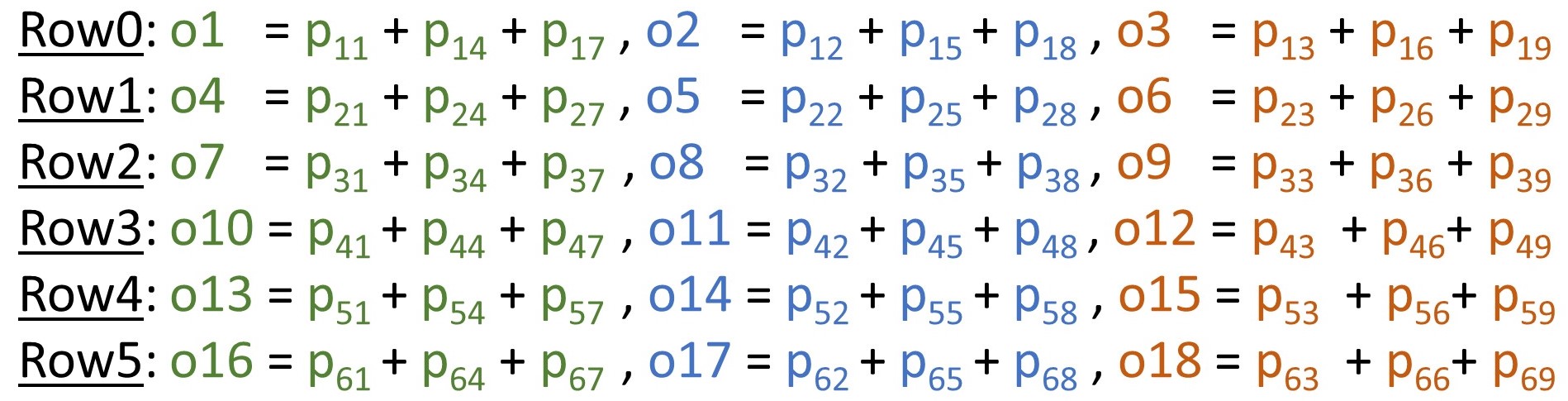}
\vspace{-3 mm}
\caption{Adder Net 0 psum Generation}
\label{addr0}
\vspace{-6 mm}
\end{figure}
\vspace{-1.5 mm}
\section{Hardware Architecture}
\vspace{-0.5 mm}
\subsection{Top-Level}
\vspace{-0.5 mm}
Figure \ref{arch} shows the top level hardware architecture of the proposed NeuroMAX CNN accelerator on Zynq-7020 SoC. The CONV core is the accelerator module containing a memory block, a state controller, PE grid, adder stages and post processing module. The memory block contains the weight, input and output SRAMs with a total cumulative size of 3.8Mb. The PE grid consists of 108 PEs arranged in 6$\times$3$\times$6 3D array. Figure \ref{arch} also shows the internal structure of the PE grid containing PE matrices numbered from 0 to 5. The PE matrices are all connected to their respective input, weight and output SRAM blocks. Each PE matrix processes independent channels in parallel for standard and separable convolutions for maximizing the throughput. The outputs from the PE matrices are provided to their respective adder nets within the PE grid. A total of six adder net 0s are present corresponding to six PE matrices. The configuration of these adder nets remain constant regardless of the type of convolution used or the filter size. The output from the adder net 0 is provided to six configurable two-stage adders whose input connections change based on the filter size and the convolution type. The first adder stage is referred to as adder net 1 and the second stage is the channel accumulation stage. \par
To perform a convolution operation, a tile of log quantized input fmap and weight data is loaded from the off-chip DDR memory into the SRAMs in the CONV core by AXI DMA and interconnect. The processor also sends the parameter information containing the values for filter size, input width, input height, output width, output height and total channels to the state controller inside the CONV core. The state controller modifies the configurable adders and determines the dataflow to be used for convolution operation. The linear convolution outputs are sent to the post processing block which performs ReLU operation and quantizes the results back into log values using pre-computed log table. These output log values are loaded into the output SRAMs and sent back to the off-chip DDR memory to be used for processing the next layer. No intermediate outputs or partial sums are stored in the DDR memory and all the intermediate processing is done within the CONV core to minimize the off-chip traffic.

\vspace{-2.5 mm}
\subsection{PE Matrix}
\vspace{-1 mm}
Figure \ref{PEmatrix} shows the hierarchical design of a single PE matrix (PE matrix 0) from left to right in a bottom up view. Each PE receives a 1D vector of weight values and one input value. It should be noted that both the input (\textit{i0'}) and the weight values (\textit{w0\textsubscript{0-2}'}) are log quantized. The output vectors from PEs are provided to the adder net 0 which generates 18 psums (o1-o18). This adder net works by summing the same color coded values generated within a row of PEs as shown in Figure \ref{addr0}.\par
Figure \ref{PEmatrix}(b) shows the internal structure of a single PE element (PE\textsubscript{0\textunderscore0\textunderscore0}). There are three compute cores or threads, each processing a single weight data and the input value, and in turn, produce three outputs (p\textsubscript{11},p\textsubscript{12},p\textsubscript{13}). The lowest level of the PE matrix is a thread within an individual PE as shown in Figure \ref{PEmatrix}(a). Basic log-based multiplication operation is performed in a single thread. Assuming we have two log quantized values, \textit{w\textsubscript{q}'} and \textit{a\textsubscript{q}'}, representing the original weight (\textit{w\textsubscript{q}}) and the activation input (\textit{a\textsubscript{q}}) respectively, the multiplication of these values in log domain can be carried out as:
\begin{equation}
    w_q a_q = sign(w_q)\cdot 2^{{g_{q}}^{'}}
    \label{exp}
\vspace{-1 mm}
\end{equation}
where,
\vspace{-2 mm}
\begin{equation}
\vspace{-2 mm}
    g_{q}^{'} = w_{q}^{'} + a_{q}^{'}
\end{equation}

\cite{logquant2} showed a method to implement the exponential in equation \eqref{exp} in hardware by decomposing the exponent into its integer and fractional part as:
\begin{equation}
\vspace{-1 mm}
    w_q a_q = sign(w_q)\cdot 2^{INT({g_{q}}^{'})}\cdot 2^{FRAC({g_{q}}^{'})}
    \label{exp2}
\vspace{1 mm}
\end{equation}

The integer part $2^{INT({g_{q}}^{'})}$ can be implemented by a shift operation, whereas, the fractional part can be pre-computed and stored within the thread. The total number of fractional computations depends on the total number of fractional bits (n) used. In our case, we have $n=1$ and thus store $2^n=2$ values in the thread memory. Equation \eqref{exp2} can now be rewritten as:
%
\begin{equation}
\vspace{0.5 mm}
    w_q a_q = sign(w_q)\cdot (LUT(FRAC({g_{q}}^{'}))>> \neg INT(({g_{q}}^{'})))
    \label{finaleq}
    \vspace{0.5 mm}
\end{equation}


The hardware implementation of equation \eqref{finaleq} is shown in Figure \ref{PEmatrix}(a). Since weights can have negative values, which is not accounted for in the log computations, we use an additional bit to represent the log weight data with the most significant bit $w_q'[6]$ representing the sign of the weight before quantization. This is not required for the input fmap values since most modern CNNs use ReLU activations which eliminate the negative outputs. \par

\begin{figure}
\includegraphics[width=0.40\textwidth,keepaspectratio]{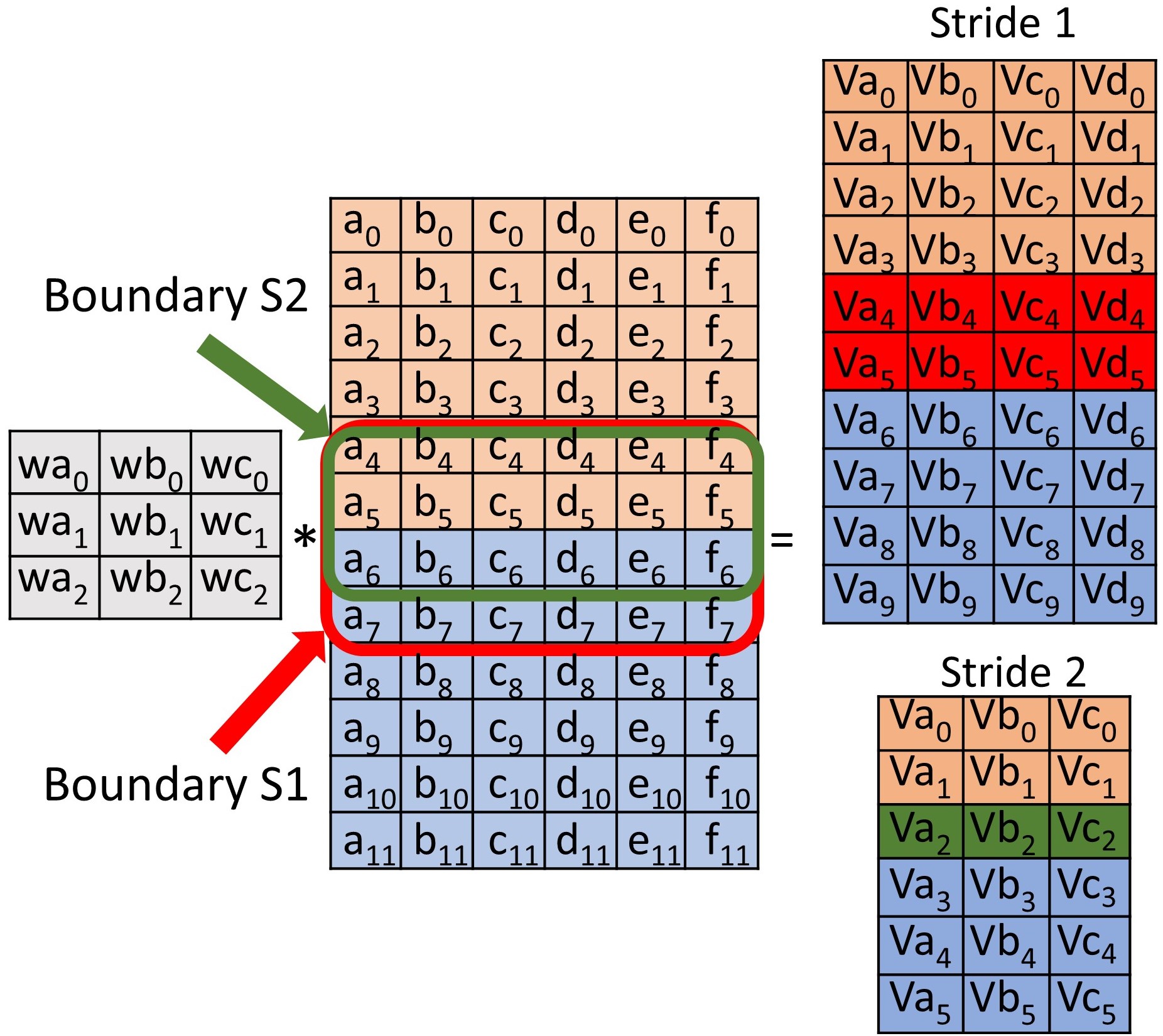}
\vspace{-2 mm}
\caption{A Convolution Example with 12$\times$6 Input, 3$\times$3 Filter for Stride 1 and 2}
\label{3_3_conv}
\vspace{-3 mm}
\end{figure}

\begin{figure}
\vspace{-2 mm}
\includegraphics[width=0.45\textwidth,keepaspectratio]{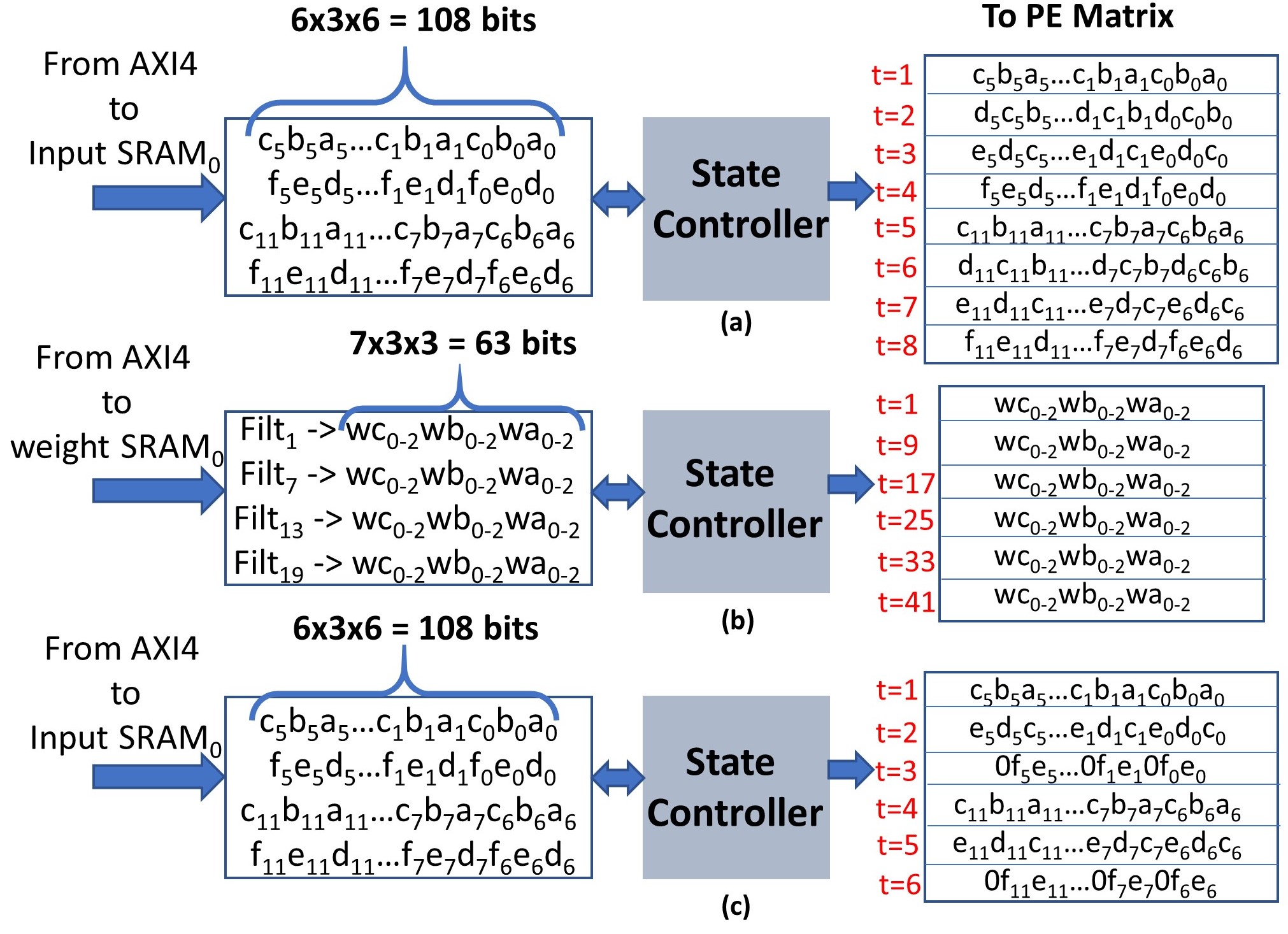}
\vspace{-5 mm}
\caption{State Controller Operation (a) Input, Stride 1 (b) Filter Weights (c) Input, Stride 2}
\label{state_controller}
\vspace{-7 mm}
\end{figure}

\begin{figure}[t]
\includegraphics[width=0.45\textwidth,keepaspectratio]{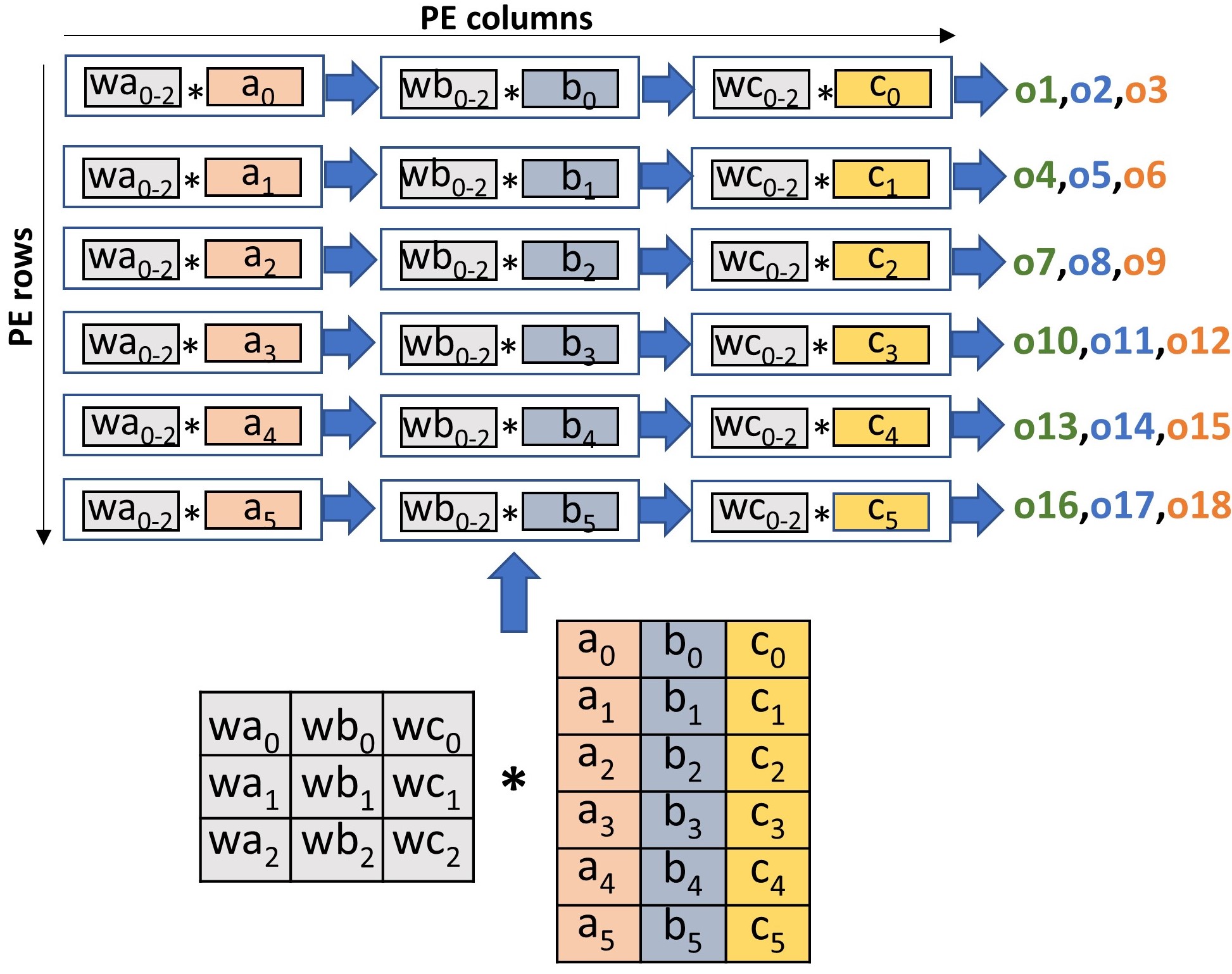}
\vspace{-4.5 mm}
\caption{2D Weight Broadcast Dataflow}
\label{DF_new1}
\vspace{-6 mm}
\end{figure}

\vspace{-1 mm}
\section{Data Flow and Processing}
The main idea behind designing an efficient dataflow is to minimize the data movement to/from the costly off-chip DDR memory. One MAC operation typically requires three memory reads corresponding to weight, ifmap, psum and one memory write corresponding to the updated psum. A neural net like AlexNet, with 724M MACs, will need $\approx$3000M DDR memory accesses. Many efficient dataflows have been presented in literature to minimize this data movement. Some of these include \textit{output stationary}, \textit{weight stationary} and, \textit{row stationary} \cite{DNNtutorial}. Since convolution operation requires the reuse of filter weights, input and psums in successive operations, the dataflows are designed to optimize the re-usability without accessing the DDR memory. We introduce a 2D weight broadcast dataflow for maximizing the re-usability of the weights, input and psums.
\begin{figure*}
\centering
\includegraphics[width=0.9\textwidth,keepaspectratio]{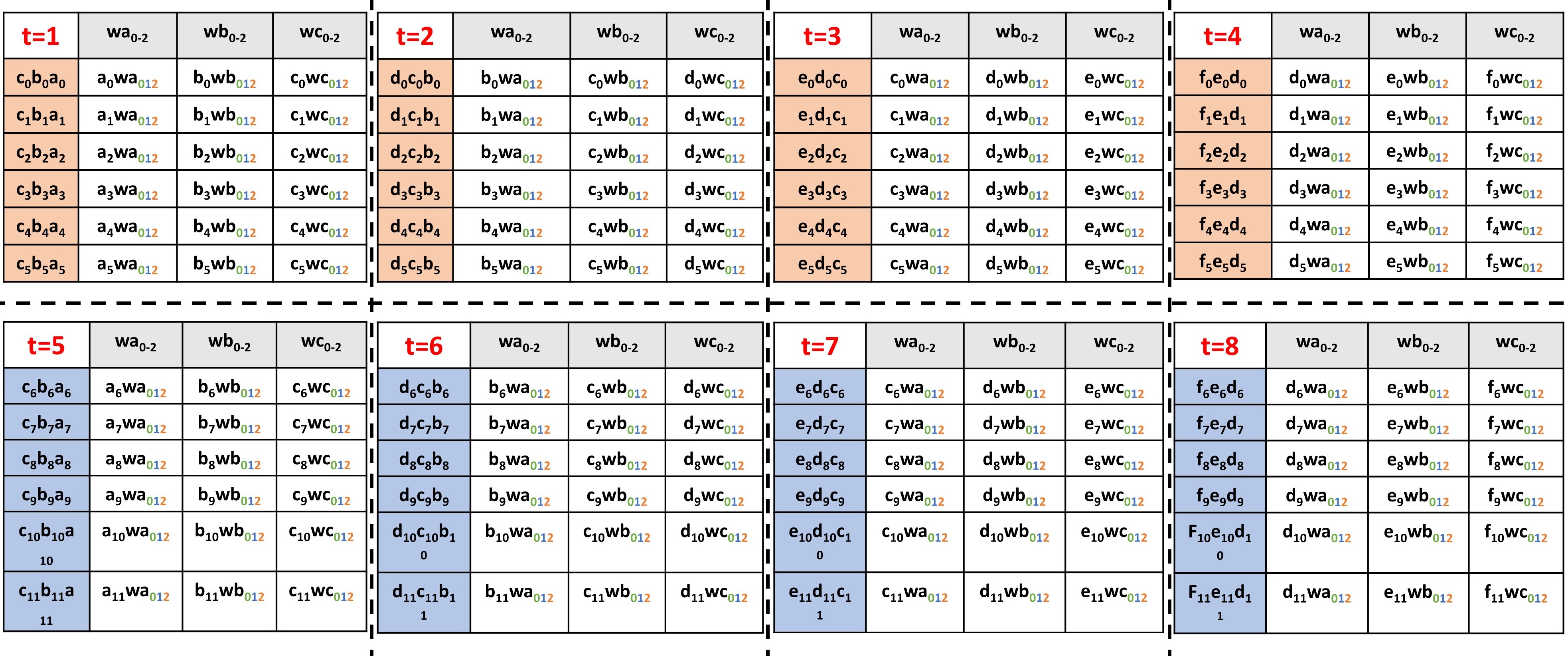}
\vspace{-4 mm}
\caption{Dataflow Chart for 3$\times$3 Stride 1 Convolution in Figure~\ref{3_3_conv}}
\label{dataflow}
\vspace{-3 mm}
\end{figure*}

\begin{figure}
\includegraphics[width=0.45\textwidth,keepaspectratio]{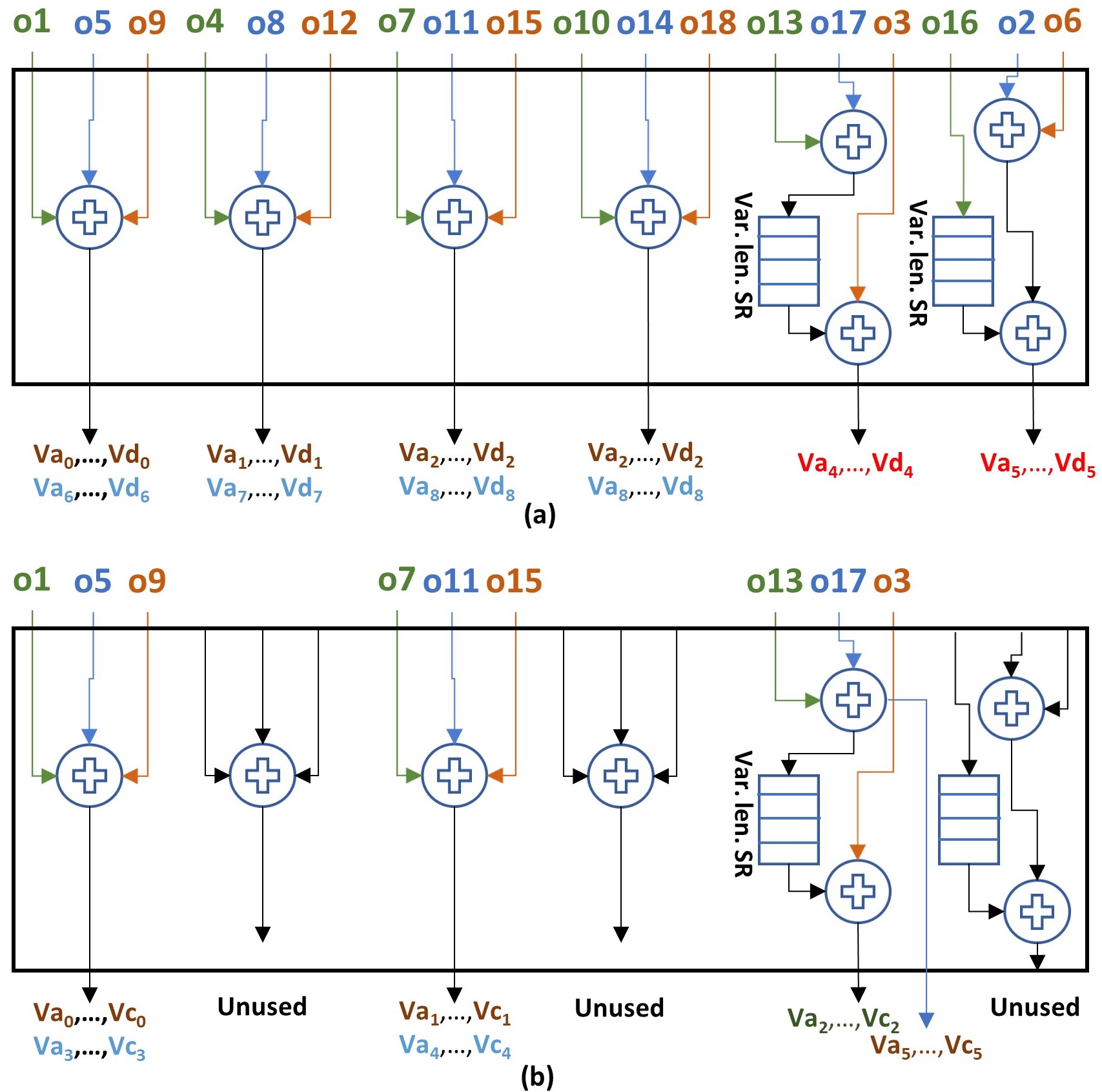}
\vspace{-4 mm}
\caption{Adder Net 1 Configuration (a) Stride 1 (b) Stride 2}
\label{addr1}
\vspace{-6 mm}
\end{figure}

\vspace{-2 mm}
\subsection{3 x 3 Convolution}
Figure \ref{3_3_conv} shows a $3\times3$ convolution example. Here, a $12\times6$ input is convolved with a $3\times3$ filter to produce a $10\times4$ output for stride 1 and a $6\times3$ output for stride 2. A total of 108 bits, corresponding to the $6\times3$ input tile, are received from the AXI4 interconnect and stored in the input SRAM. This input tile is modified by the state controller and provided to the PE matrix in a row shifted pattern as shown in Figure \ref{state_controller}(a) and (c) for stride 1 and stride 2, respectively. We also acquire a 2D weight array and broadcast it to the PE matrix as shown in Figure \ref{state_controller}(b). Figure \ref{DF_new1} shows the dataflow and the operation of the PE matrix for the first 6x3 input tile and the weight matrix at time stamp t = 1. The entire input tile and the 2D weight array is loaded into the PE matrix simultaneously. Because of the multi-threaded structure of PEs, each PE performs three multiplication operations using three threads and the outputs are row-wise summed to generate the psums (o1-o18) using adder net 0 (Figure \ref{addr0}). The dataflow chart and the processing of the entire $12\times6$ input is shown in Figure \ref{dataflow}. The output a\textsubscript{0}wa\textsubscript{012}, in Figure \ref{dataflow}, represents the three outputs a\textsubscript{0}wa\textsubscript{0}, a\textsubscript{0}wa\textsubscript{1}, a\textsubscript{0}wa\textsubscript{2} generated by three threads within a PE. The adder net 0 computes the partial sum outputs (o1-o18) the same way as shown in Figure \ref{addr0}, where p\textsubscript{11} = a\textsubscript{0}wa\textsubscript{0}, p\textsubscript{14} = b\textsubscript{0}wb\textsubscript{0} and p\textsubscript{17} = c\textsubscript{0}wc\textsubscript{0} are the same colored (green) outputs along the row.\par
The dark red outputs (row 5 and 6) in Figure \ref{3_3_conv} for stride 1 and green outputs (row 3) for stride 2 represent the boundary outputs. The boundary condition occurs when the filter overlaps two different column-wise input tile sectors. For clarity, we assume that the first input tile at t = 1 is processed by the PE matrix. This corresponds to the first six rows and the first three columns of the input. The PE matrix will process the last row-wise input tile at t = 4 which corresponds to the first six rows and the last three columns of the input as shown in Figure \ref{state_controller}(a). The input tile will then jump to the next column-wise 6x3 input tile which corresponds to the last six rows and the first three columns at t = 5 as shown in Figure \ref{state_controller}(a). However, it can be seen that the row 5 and 6 in the output are dependent on the overlapping results from the two concurrent column-wise input tile sectors (e.g. at t = 1 and t = 5, t = 2 and t = 6 and so on). To resolve this, the three dependent psums (o13, o17 and o16), generated from row 5 and row 6, of first column-wise tile sector of the 12x6 input, are passed through a variable length shift register with the maximum length equal to the width of the input. These psums are subsequently utilized when the next column-wise 6x3 input tile (a\textsubscript{6} to a\textsubscript{11}) is being processed. Thus, the rows 1 to 4 in the output are generated during the time intervals t = 1 to t = 4, whereas the rows 5 to 10 are generated during the time interval t = 5 to t= 8.
\par
The output in Figure \ref{3_3_conv}, for stride 1, is generated by alternate colored, column-wise summation of the psums in the adder net 1 as shown in Figure \ref{addr1}(a). Figure \ref{addr1}(b) shows the output generation for stride 2 case. The shift registers (VAR Len SR) for generating the boundary outputs are also shown in Figure \ref{addr1}. It can be observed that because of the optimized dataflow, only 2 out of 18 or 11\% psums require local storage as opposed to >50\% psums requiring storage (local or off-chip) in previously proposed dataflows. The throughput for the above example is $45$ OPS/cycle (total OPS/total cycles = 360/8 = 45), which results in an $83.3\%$ overall thread utilization (45/($3\times6\times3$))$\times100$. We will simply use thread utilization as hardware utilization in this context. 

\vspace{-2 mm}
\subsection{1 x 1 Convolution}
\vspace{-1 mm}
$1\times1$ convolutions are very popular in modern CNNs. These convolutions, along with the depth-wise separable, are replacing the normal 2-D convolutions because of the less number of MAC operations \cite{mobnetv1}. The $1\times1$ CONV operation convolves $1\times1\times C \times P$ filters with a $M\times N\times C$ input to produce $M\times N\times P$ outputs. Here, C is the number of channels, P is the number of filters, M is the input width and N is the input height. \par

\begin{figure}
\includegraphics[width=0.47\textwidth,keepaspectratio]{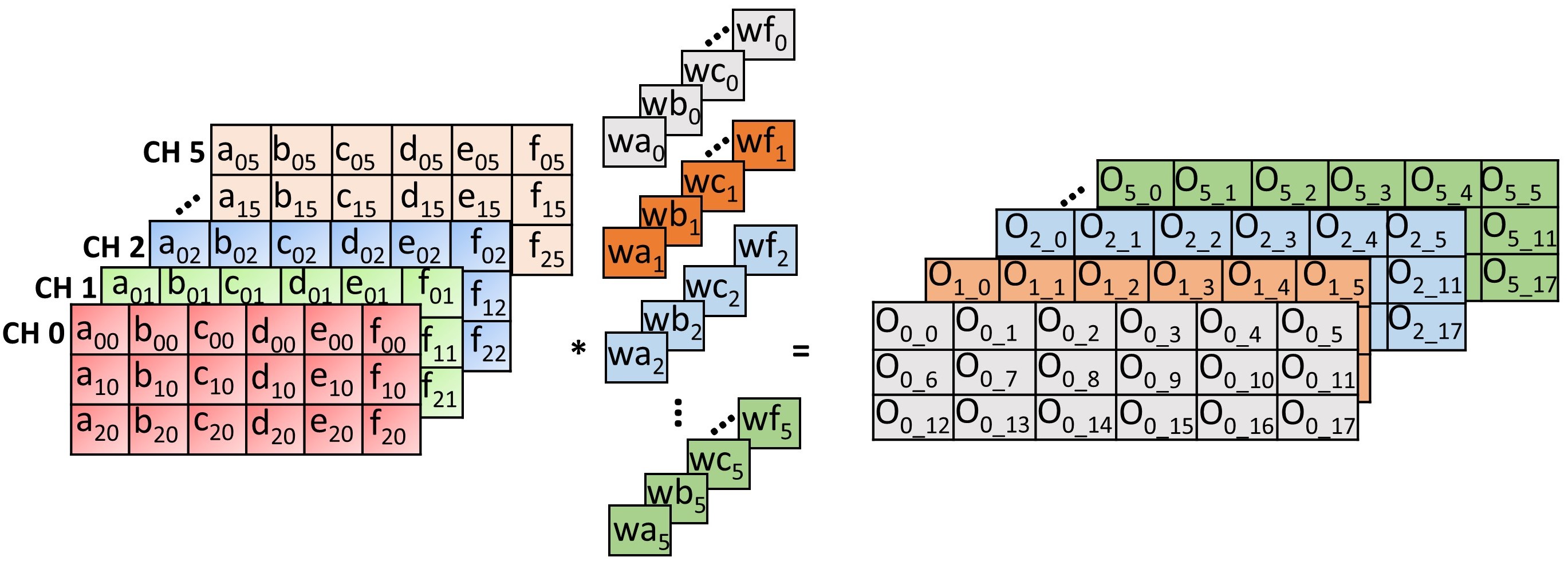}
\vspace{-5 mm}
\caption{$1\times1$ Convolution Example}
\label{pt_conv}
\vspace{-1 mm}
\end{figure}

\begin{figure}
\includegraphics[width=0.45\textwidth,keepaspectratio]{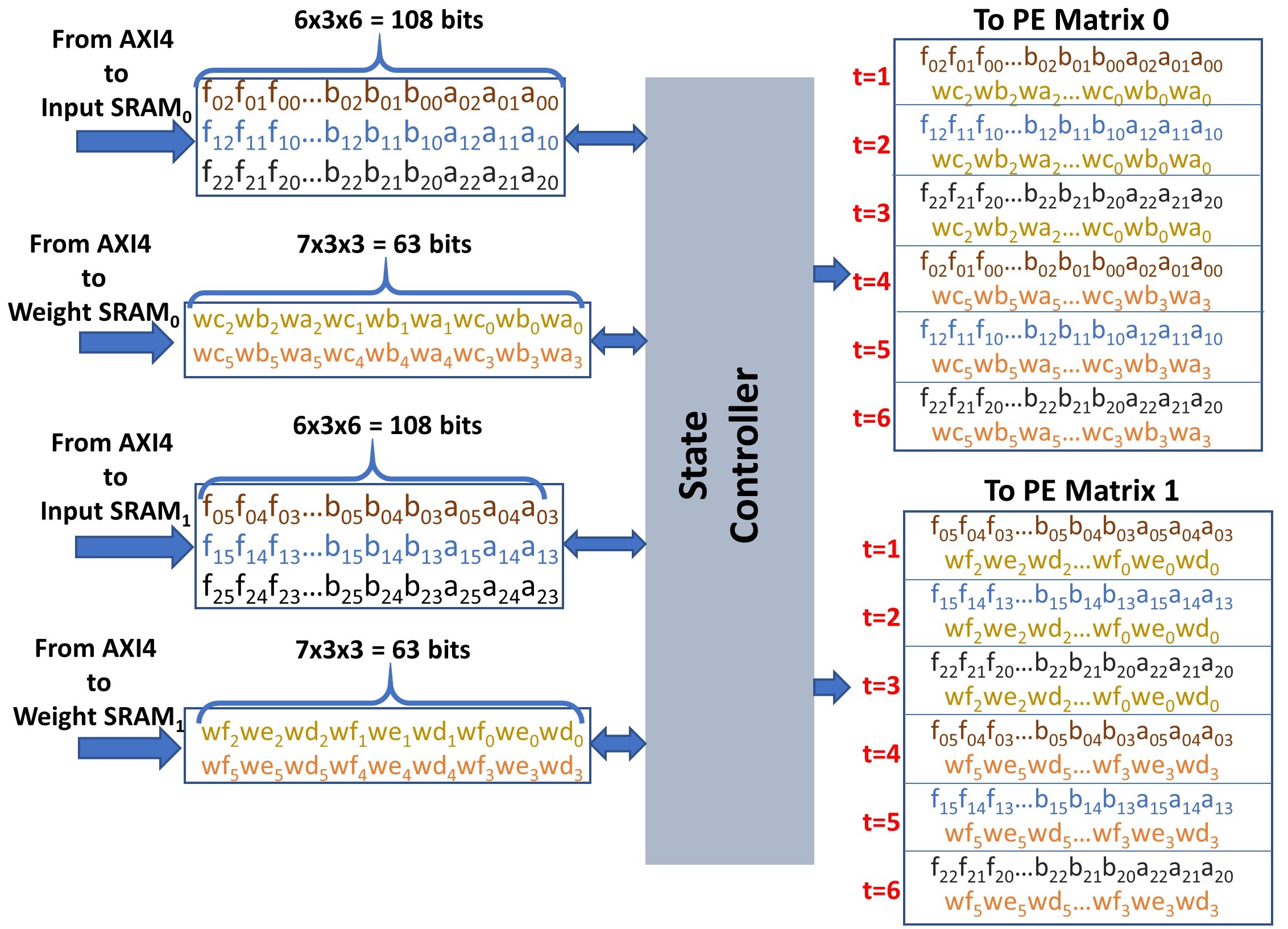}
\vspace{-4.5 mm}
\caption{State Controller Load Operation}
\label{pt_conv_SC}
\vspace{-4 mm}
\end{figure}

\begin{figure}[t]
\includegraphics[width=0.45\textwidth,keepaspectratio]{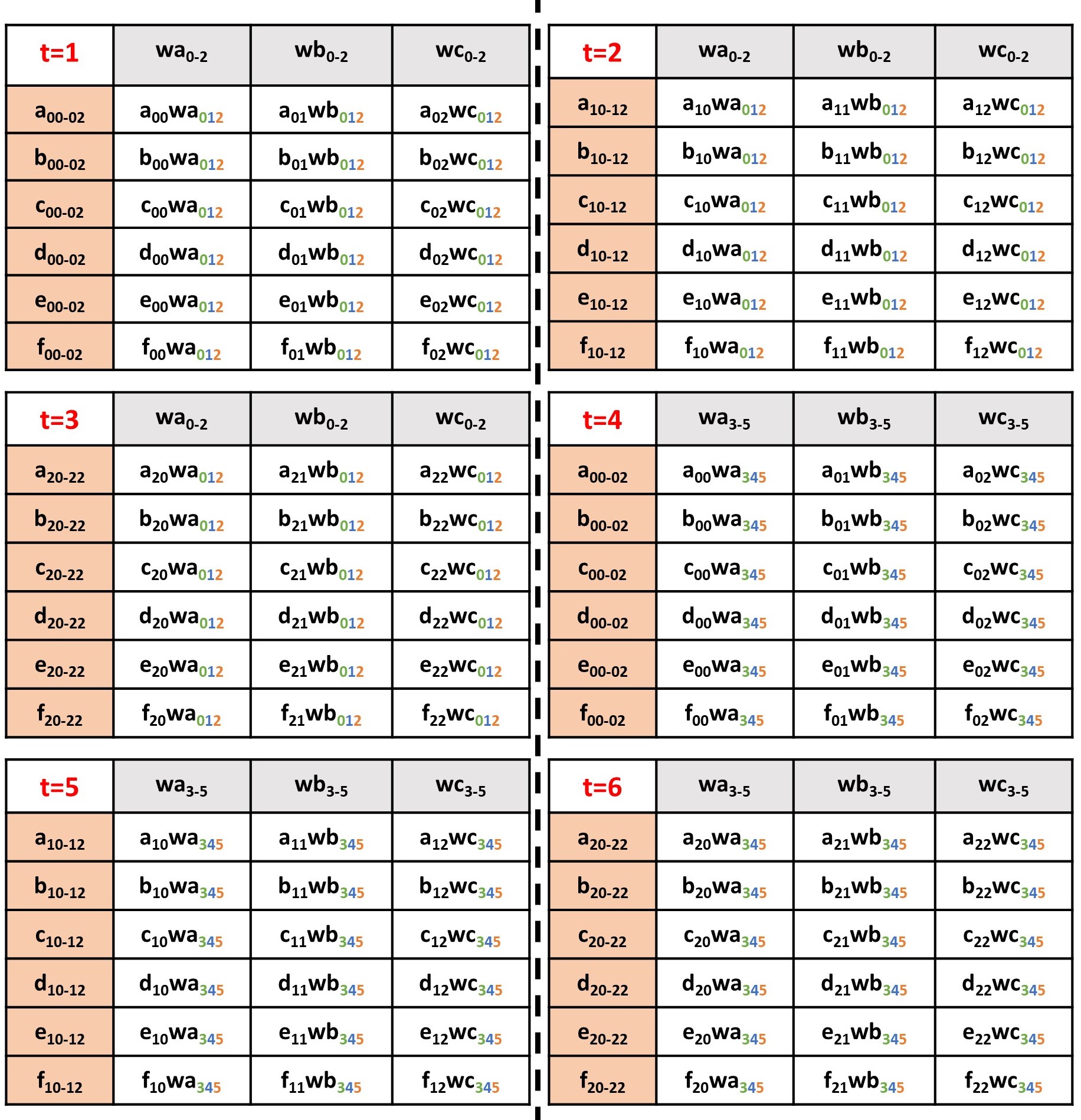}
\vspace{-4 mm}
\caption{Dataflow Chart for 1$\times$1 Convolution in Figure~\ref{pt_conv} }
\label{pt_conv_DF}
\vspace{-5 mm}
\end{figure}

\begin{figure}[t]
\vspace{-2 mm}
\includegraphics[width=0.47\textwidth,keepaspectratio]{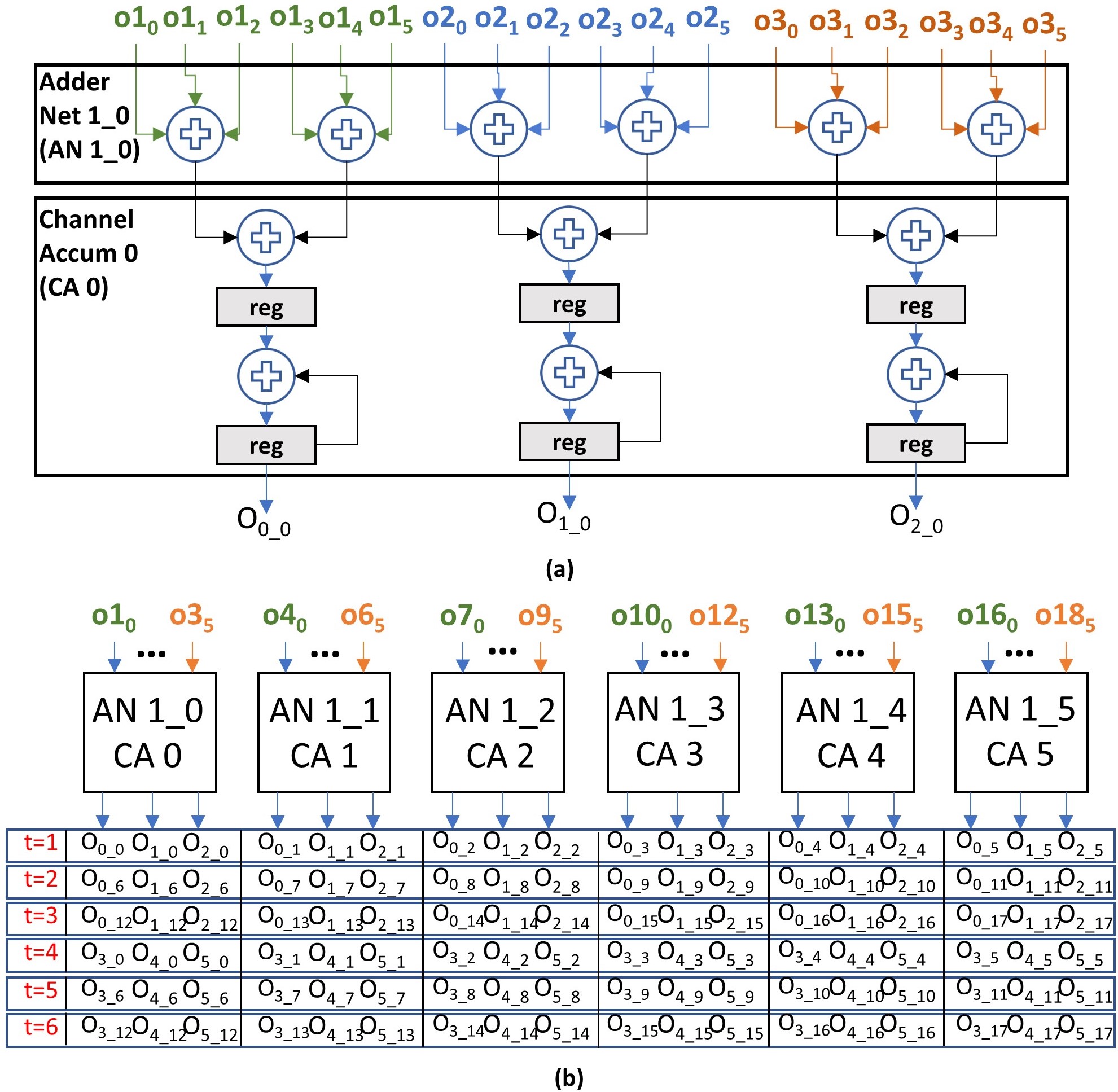}
\vspace{-5 mm}
\caption{(a) Channel-wise accumulation for PE matrix 0 (b) All channel-wise accumulations }
\label{ch_accum}
\vspace{-4 mm}
\end{figure}
\begin{figure}
\includegraphics[width=0.45\textwidth,keepaspectratio]{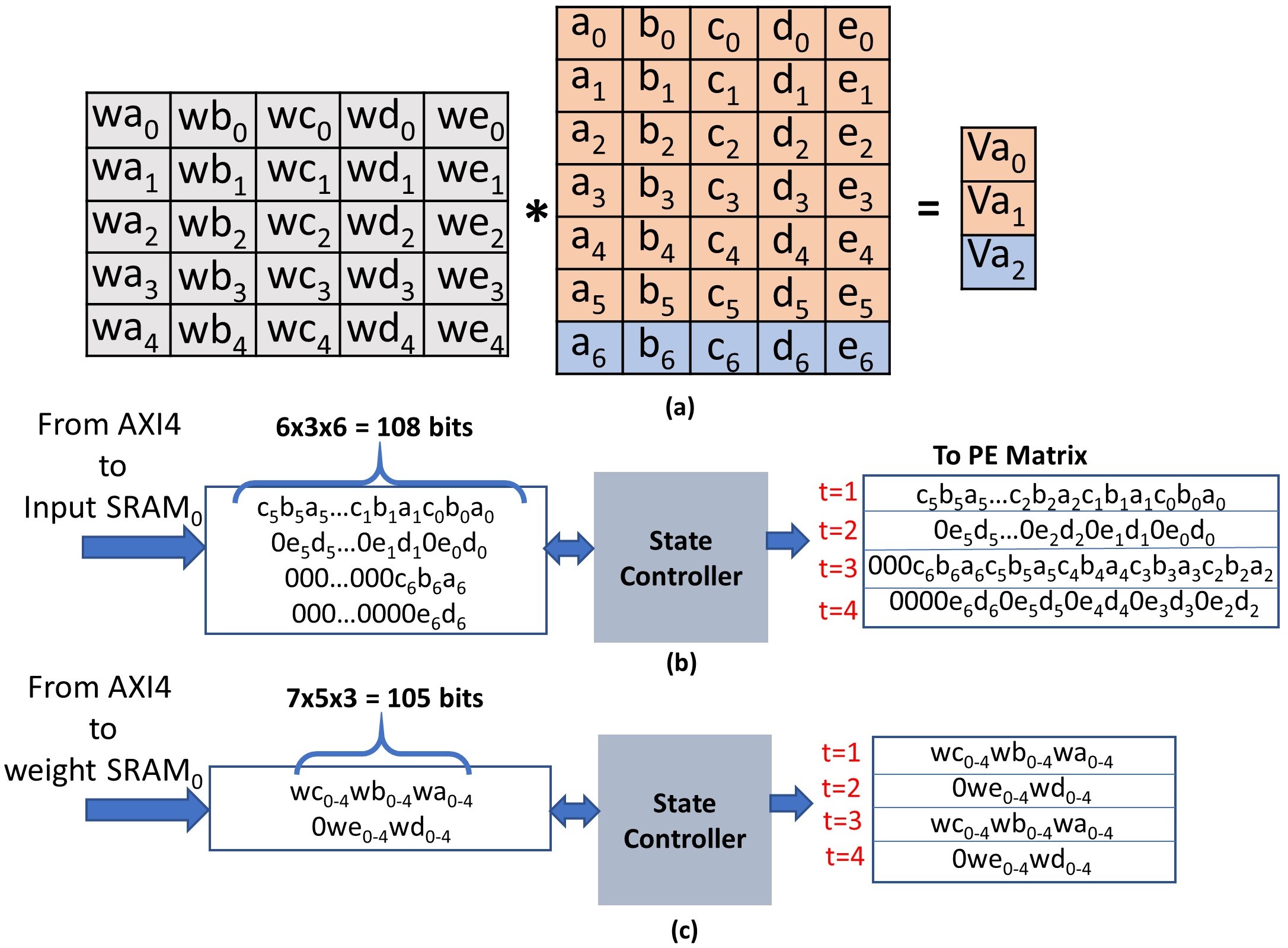}
\vspace{-5 mm}
\caption{(a) $5\times5$ Convolution Example (b) Input Load Operation (c) Weight Load Operation}
\label{5_5_conv}
\vspace{-6 mm}
\end{figure}

Figure~\ref{pt_conv} shows a $1\times1$ CONV example where a $3 \times 6 \times 6$ input is convolved with 6, $1 \times 1 \times 6$ filters to produce a $3\times6\times6$ output. Since this convolution generates the psums by channel accumulation, the outputs from the multiple PE matrices are utilized. For the example in Figure \ref{pt_conv}, the state controller data scheduling for PE matrix 0 and 1 is shown in Figure \ref{pt_conv_SC}. It can be seen that the first three channels of the input are convolved with the first three channels of all the filters in PE matrix 0, whereas, the last three channels of the input are convolved with the last three channels of all the filters in PE matrix 1. The time stamps during specific processing of input and weights in PE matrices are also shown in Figure \ref{pt_conv_SC}. It should be noted that for an input with more channels, the rest of the PE matrices will also be used. Thus, by using the dataflow in Figure \ref{pt_conv_SC}, the architecture can process 18 channels concurrently by using the 6 PE matrices, with each PE matrix processing 3 input and filter channels.\par
The dataflow chart for the PE matrix 0 for the example in Figure \ref{pt_conv} is shown in Figure \ref{pt_conv_DF}. The same dataflow chart can also be generated for the PE matrix 1. As mentioned earlier, the psums in $1\times1$ convolution are calculated using channel-wise accumulation. The eighteen outputs (o1-o18) generated by the individual PE matrices are summed in their respective adder net 1s. The input connections for adder net 1 (AN 1\textunderscore0) and the channel accumulator (CA 0) of PE matrix 0 are shown in Figure \ref{ch_accum}(a). Here, o1\textsubscript{0} is the psum output from the PE matrix 0 and o1\textsubscript{5} is the psum output from the PE matrix 5. Since the example in Figure \ref{pt_conv} is small, it only requires the first two PE matrices and their outputs, that is, only o1\textsubscript{0}-18\textsubscript{0} and o1\textsubscript{1}-18\textsubscript{1} are active. The output in Figure \ref{pt_conv} is generated by using all six adder net 1s and the channel accumulators as shown in Figure \ref{ch_accum}(b). The throughput for the above example is $108$ OPS/cycle (total OPS/total cycles = ($6\times6\times3\times6/6 = 108$), which results in a $100\%$ overall thread utilization (108/($3\times6\times3\times2$))$\times100$.

\vspace{-2.5 mm}
\subsection{Higher Order Convolutions}
\begin{figure}
\includegraphics[width=0.45\textwidth,keepaspectratio]{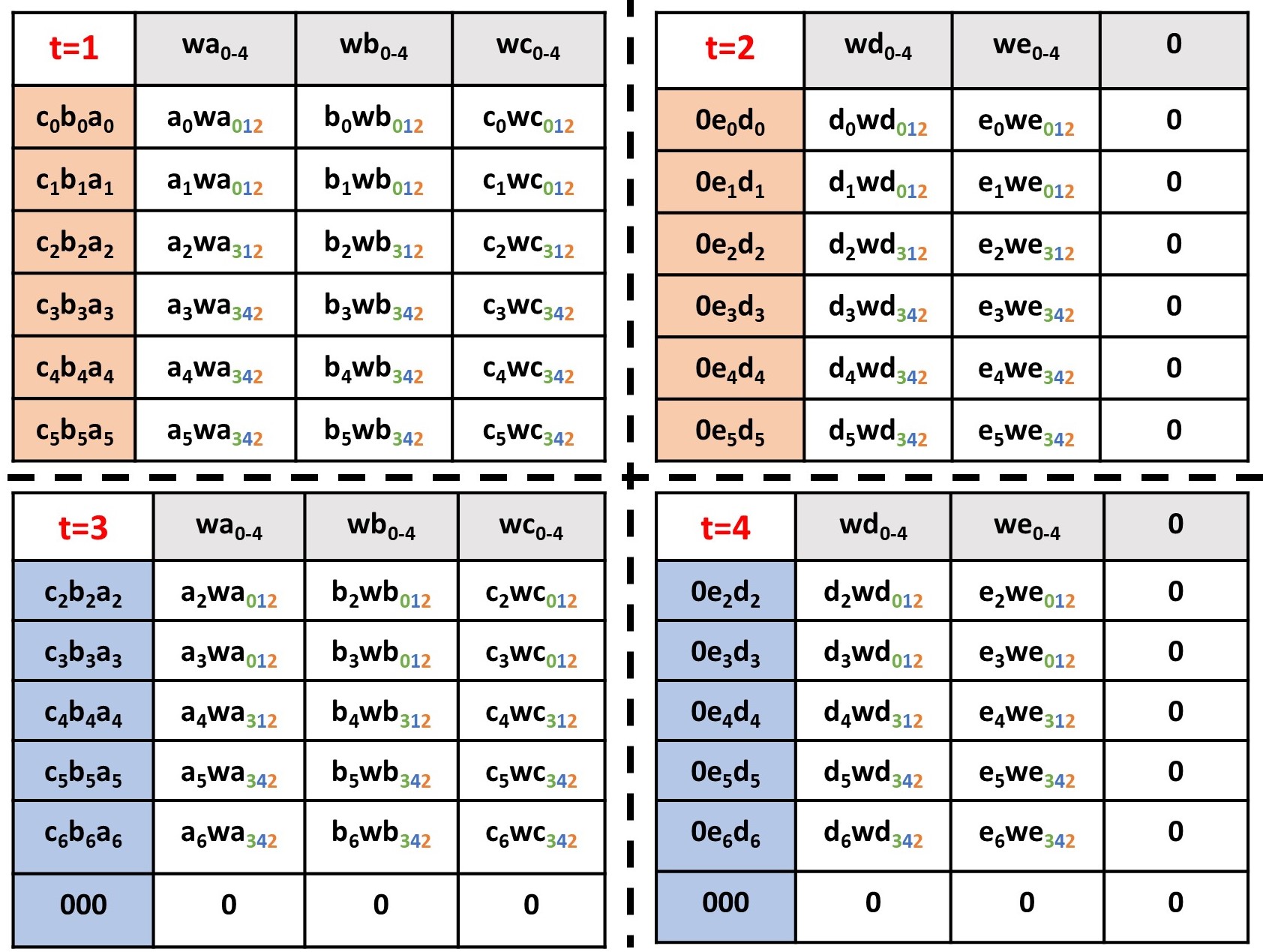}
\vspace{-4 mm}
\caption{Dataflow Chart for $5\times5$ Convolution in Figure~\ref{5_5_conv}(a)}
\label{5_5_conv_DF}
\vspace{-2 mm}
\end{figure}

\begin{figure}
\vspace{-1.5 mm}
\includegraphics[width=0.45\textwidth,keepaspectratio]{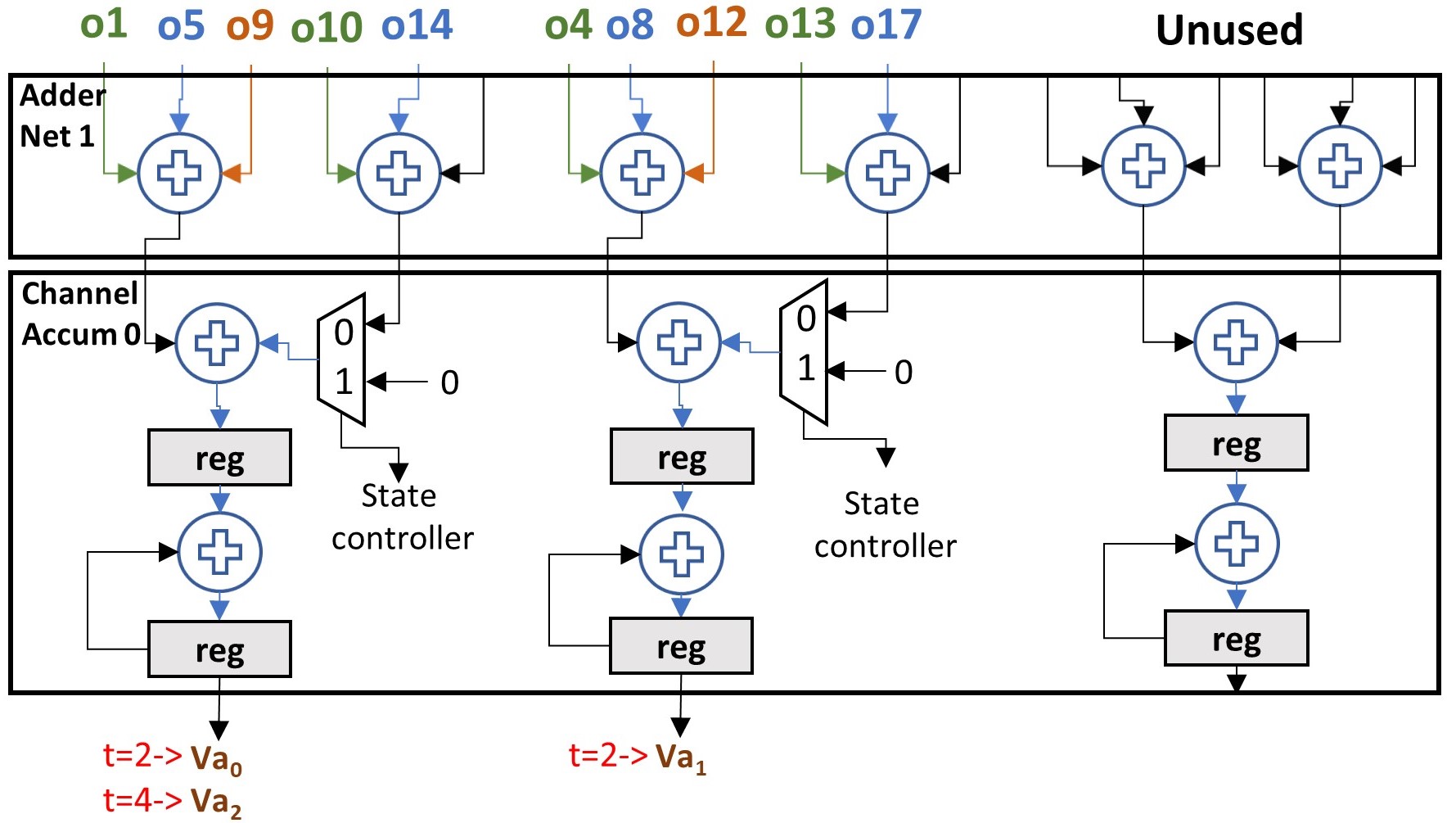}
\vspace{-4 mm}
\caption{Adder Configuration for $5\times5$ Convolution}
\label{5_5_conv_addr}
\vspace{-6 mm}
\end{figure}

\vspace{-1 mm}
The proposed NeuroMAX accelerator is designed to optimize $3\times3$ and $1\times1$ convolutions. It can, however, also be used to accelerate larger kernel sizes. \cite{decompose} proposed a kernel decomposition method such that an additional support for $4\times4$ and $5\times5$ filter is needed to implement any filter size. Figure \ref{5_5_conv}(a) gives an example of $5\times5$ convolution. As the size of the PE matrix is $6\times3$, a filter of width greater than 3 and height greater than 6 needs multiple cycles to calculate the output value. This can be seen in Figure \ref{5_5_conv}(b) and (c) where the last two columns of the input matrix and the weight matrix are loaded at time stamp t = 2. Figure \ref{5_5_conv_DF} shows the dataflow chart which accounts for this configuration. The generated psums (o1-o18) are provided to the adder net 1 as shown in Figure \ref{5_5_conv_addr}. For this convolution, the output values are calculated as:
\vspace{0.5 mm}
\begin{equation}
   Va_0,Va_2 = ((o1 + o5 + o9) + (o10 + o14))_{old} + (o1 + o5 + o9)_{new}
    \label{val1}
\end{equation}
\begin{equation}
   Va_1 = ((o4 + o8 + o12) + (o13 + o17))_{old} + (o4 + o8 + o12)_{new}
    \label{val2}
\end{equation}

In equations \eqref{val1} and \eqref{val2}, the \textit{old} value corresponds to the convolution output from the first three columns of the input and the weight matrix at $t = 1$, whereas, the \textit{new} value corresponds to the last two columns at $t = 2$. The adder net 1 and the channel accumulator configuration for this convolution is shown in Figure \ref{5_5_conv_addr}. A similar configuration and dataflow chart is used for implementing a $4\times4$ convolution. In addition to this, the CONV core can also perform pooling operation by choosing the appropriate stride and kernel.

\vspace{-2 mm}
\section{Implementation and Results}
\begin{figure}[h]
\textbf{\vspace{-3 mm}}
\includegraphics[width=0.47\textwidth,keepaspectratio]{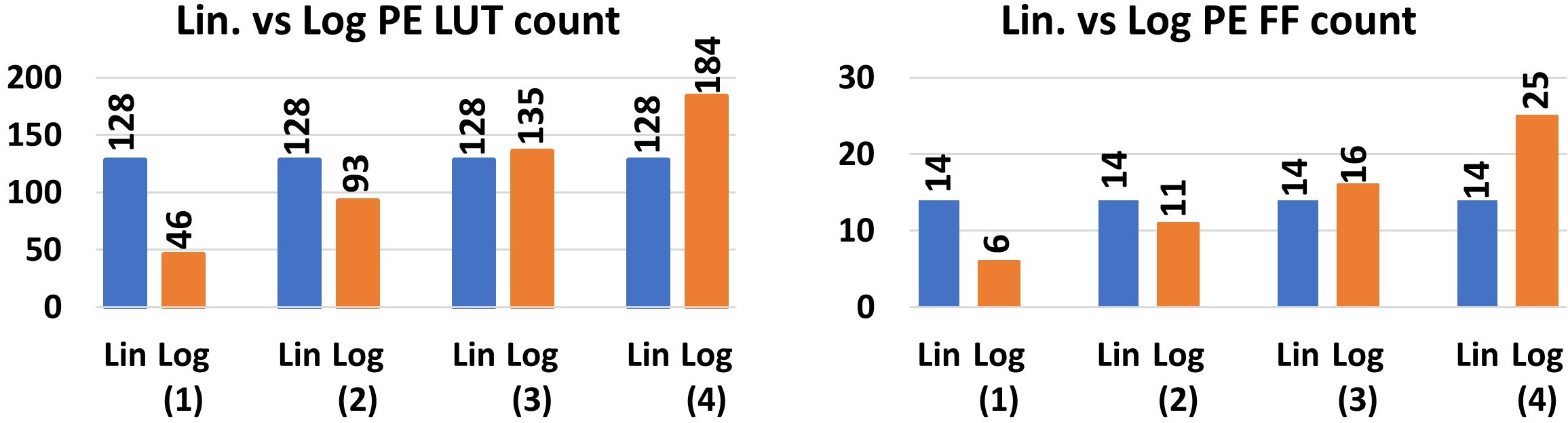}
\vspace{-3 mm}
\caption{Linear vs Log PE LUT and FF Cost at 16-bit Precision}
\label{PE_cost}
\vspace{-4 mm}
\end{figure}

This section discusses the implementation of the proposed NeuroMAX accelerator and presents the area cost, power consumption, performance, throughput, and hardware utilization results. The accelerator has been implemented on the PL side of Xilinx Zynq-7020 SoC operating at 200MHz. Figure \ref{PE_cost} shows cost comparison between our multi-threaded log PE core and an area optimized linear multiplier core with equal output bit precision and latency.
It can be seen that by choosing a thread count of 3 (shown as log (3) in Figure~\ref{PE_cost}), the LUT and FF cost is only $1.05\times$ and $1.14\times$ that of the linear PE. Thus, a total of 108 linear PEs would be equivalent, in cost, to $\approx$122 multi-threaded log PEs. For fairness, we will use the cost adjusted PE number for performance comparison. 

\begin{table}[t]
  \caption{Resource Utilization}
  \label{table1}
  \vspace{-3 mm}
\addtolength{\tabcolsep}{3pt}
  \begin{tabular}{cccl}
    \toprule
    Property & \ Accelerator & \  Utilization\\
    \midrule
    \#LUTs & 20680 & $38\%$\\
    \#FFs & 17207 & $16\%$ \\
    \#36kB BRAMs & 108 & $77\%$ \\
     Power  & 2.727 W & NA \\
    \bottomrule
  \end{tabular}
  \vspace{-1 mm}
\end{table}

\begin{table*}
\centering
  \vspace{-1 mm}
  \caption{Comparison of NeuroMAX with Previous Designs}
  \label{table2}
  \vspace{-4 mm}
\addtolength{\tabcolsep}{3pt}
  \begin{tabular}{cccccccl}
    \toprule
    Property & \ NeuroMAX & \ \cite{eyeriss}    &\ \cite{fpgaCNN1} &\ \cite{fpgaCNN2} & \ \cite{eyerissv2} &\ \cite{logquant2} & \ \cite{VWA}  \\
    \midrule
    \ Technology & Zynq-7020 SoC & 65nm  & Zynq-7100 & Arria 10 SoC & 65nm & Virtex-7 & 40nm   \\
    \ Precision(bits) & 6-bit log  & 16-bit & 32fp & 16-bit & 8-20 bits & 5-bit log & 16-bit  \\
    \ PE number & 122(adjusted) & 168  & 1926 & 1278 & 192 & 256 & 168   \\
    \ Processing clock (MHz)  & 200 & 200  & 100 & 133 & 200 & Unreported & 500  \\
    \ Peak Throughput (GOPS) & 324 & 84  & 17.11 & 170.6 & 153.6 & Unreported &  168  \\
    \ Peak Throughput/PE & 2.7(adjusted) & 0.5  & 0.008 & 0.13 & 0.8 & Unreported & 1  \\
    \ Cost (LUTs(a),gates(b)) & 20.6k(a) & 1176k(b)  & 142k(a) & 66k(a) & 2695k(b) & 29k(a) & 266k(b)  \\
    \ Power (W) & 2.72 & 0.278  & 4.083 & Unreported & 0.460 & 3.756 & 0.155  \\
    \bottomrule
  \end{tabular}
  \vspace{-3 mm}
\end{table*}

\begin{figure}
\includegraphics[width=0.40\textwidth,keepaspectratio]{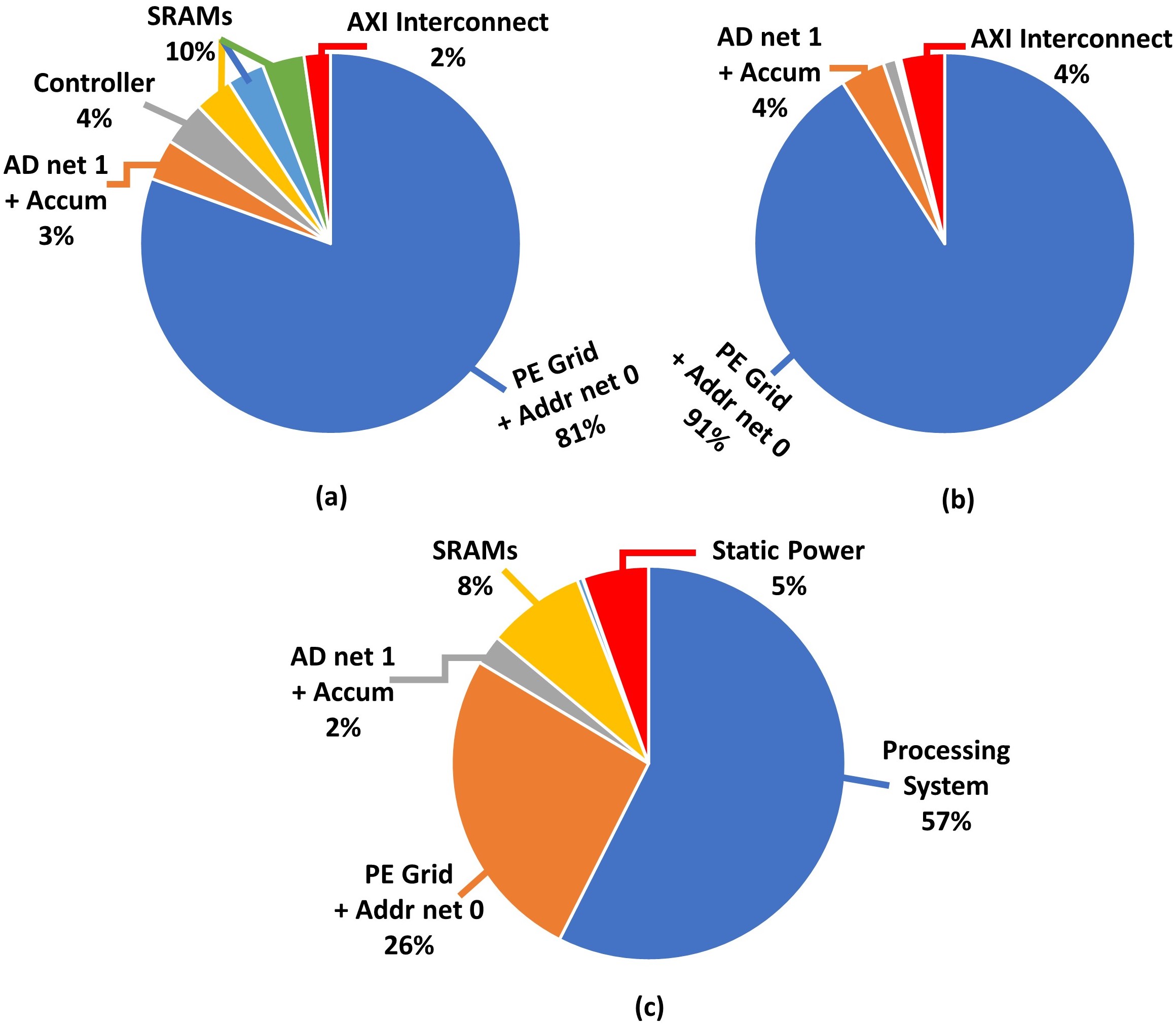}
\vspace{-4 mm}
\caption{Breakdown of  (a) LUT cost (b) FF cost (c) Power Consumption for NeuroMAX}
\label{hardwareUtil}
\vspace{-6 mm}
\end{figure}
Table \ref{table1} shows the resource utilization of the implemented accelerator core as well as the total power consumption (static + dynamic). Figure \ref{hardwareUtil} (a), (b), (c) shows the breakdown of LUT cost, FF cost and power consumption among different modules of the accelerator. The PE grid and the adder net 0 combined have the highest LUT and FF count (81\% and 91\%, respectively). The post processing block consumes negligible resources. The processing system (ARM core) dominates the power consumption (57\%), while the PE grid and adder net 0 have the second highest consumption (26\%) of the total.

\begin{figure}
\includegraphics[width=0.45\textwidth,keepaspectratio]{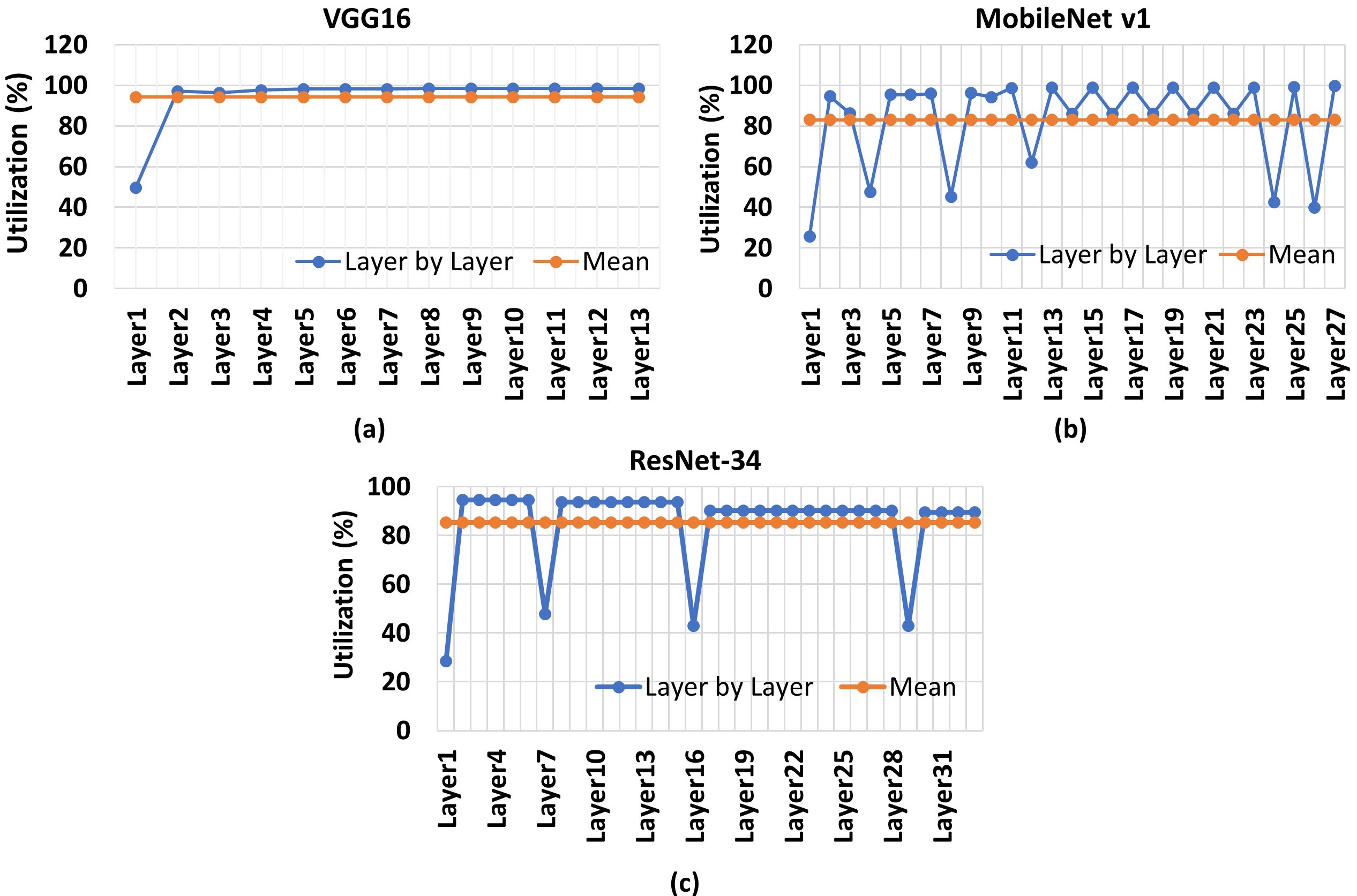}
\vspace{-4.5 mm}
\caption{Hardware Utilization of NeuroMAX for (a)VGG-16 (b) MobileNet v1 (c) ResNet-34}
\label{Util_comp}
\vspace{-3 mm}
\end{figure}

Figure \ref{Util_comp} shows layer by layer hardware utilization for various CNN architectures. We achieve an average utilization of $95\%$, $84\%$ and, $86\%$ for VGG-16, MobileNet v1 and, ResNet-34, respectively. The dip in hardware utilization in some layers of mobilenet and ResNet-34 is because of stride 2 convolutions which utilize only $50\%$ of the available PE cores. The low utilization in the first layer of VGG16 is because it only has 3 channels and since each PE matrix processes one channel, the last 3 PE matrices remain idle which gives an exact utilization of $50\%$.

\begin{figure}
\includegraphics[width=0.45\textwidth,keepaspectratio]{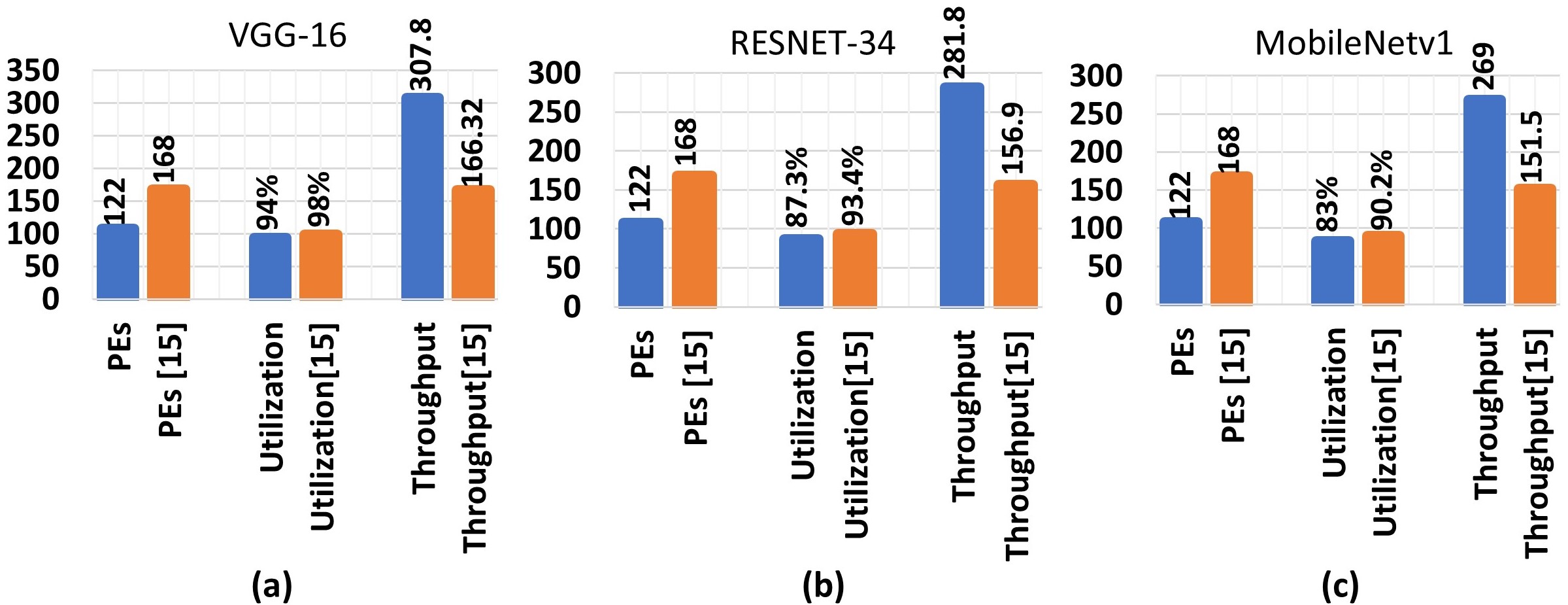}
\vspace{-3 mm}
\caption{PE count vs Utilization vs Throughput Comparison of NeuroMAX with \cite{VWA}}
\label{VWA_comp}
\vspace{-6 mm}
\end{figure}


Chang et al. \cite{VWA} recently presented an accelerator design with 1D broadcast dataflow which promises higher utilization and throughput (GOPS) then all the previous designs. We, therefore, compare our design against \cite{VWA} in Figure \ref{VWA_comp} for various CNNs. \cite{VWA} uses a total of 168 PE cores and provides a utilization of $99\%$ with throughput $166.32$ GOPS, $93.4\%$ with throughput $156.91$ GOPS  and, $90.2\%$ with throughput $151.54$ GOPS for VGG16, ResNet-34 and mobilenet, respectively. We use 122 PE cores (cost adjusted), a $28\%$ decrease from \cite{VWA}, and provide a throughput of $307.8$ GOPS, an $85\%$ increase, $281.8$ GOPS, a $79.4\%$ increase, and $268.92$ GOPS, a $77.4\%$ increase, for the three CNNs, respectively. This increase in throughput with lower PE count is attributed towards our low cost, multi-threaded PE core design and an efficient 2D dataflow. We also achieve somewhat similar hardware utilization, that is, $94\%$ for VGG16, $87.3\%$ for ResNet-34 and, $83\%$ for mobilenet. It should be noted that \cite{VWA} implements the accelerator on an ASIC, whereas, we use an FPGA, thus, an accurate comparison in LUT count, FF count and, power consumption cannot be made. It is, however, evident that the design in \cite{VWA} when ported into FPGA will have $\approx$31\% more LUTs and FFs owing to more number of PEs used. \par

\begin{table}[ht]
  \vspace{1 mm}
  \caption{VGG16 Latency Comparison}
  \label{latency}
  \vspace{-2 mm}
\addtolength{\tabcolsep}{3pt}
  \begin{tabular}{ccccl}
    \toprule
    \textbf{Layer} & \ \textbf{NeuroMAX} & \ \textbf{\cite{eyeriss}} & \ \textbf{\cite{VWA}}  \\
    \midrule
    CONV1\textunderscore1(ms) & 1.35 & 38.0& 2.57  \\
    CONV1\textunderscore2(ms) & 28.9 & 810.6 & 55.04 \\
    CONV2\textunderscore1(ms) & 14.4 & 405.3& 27.43  \\
    CONV2\textunderscore2(ms) & 29.26 & 810.8& 55.7  \\
    CONV3\textunderscore1(ms) & 14.54 & 204& 27.7  \\
    CONV3\textunderscore2(ms) & 28.6 & 408.1& 54.5  \\
    CONV3\textunderscore3(ms) & 28.7 & 408.1& 54.6  \\
    CONV4\textunderscore1(ms) & 14.4  & 105.1& 27.42\\
    CONV4\textunderscore2(ms) & 29 & 210.0& 55.23   \\
    CONV4\textunderscore3(ms) & 29.5 & 210.0& 56.19 \\
    CONV5\textunderscore1(ms) & 7.24 & 48.3& 13.79  \\
    CONV5\textunderscore2(ms) & 7.23 & 48.5& 13.77  \\
    CONV5\textunderscore3(ms) & 7.11 & 48.5& 13.54   \\
    \textbf{Total(ms)} & \textbf{240.23} & \textbf{3755.3} & \textbf{457.5} \\
    \bottomrule
  \end{tabular}
  \vspace{-2 mm}
\end{table}

Table \ref{table2} shows the comparison of our accelerator with previous state of the art ASIC and FPGA designs. We see an improved performance in terms of PE number, peak throughput and peak throughput/PE ratio. Only \cite{VWA} has a peak throughput/PE ratio equal to unity with average around 0.85. Our peak throughput/PE is 3 with average around 2.7 after cost adjustment. The power comparison reveals that the FPGA-based designs inherently consume more power compared to ASICs. We can, however, see that NeuroMAX consumes significantly less power and has lower cost in terms of LUT count compared to other FPGA designs.

Table \ref{latency} gives a layer-by-layer  processing latency comparison for VGG16. Both \cite{eyeriss} and \cite{VWA} benchmark the latency of their accelerators on this CNN, therefore, we also evaluate and compare NeuroMAX's performance on VGG16. It should be noted however that \cite{VWA} uses 500MHz processing clock in their design. For fair comparison, we make suitable adjustments in their reported values. Our proposed NeuroMAX accelerator has $93\%$ and, $47\%$ decrease in latency, when compared to \cite{eyeriss} and \cite{VWA}, respectively, at 200 MHz clock.

\vspace{-3 mm}
\section{Conclusion}
\vspace{-1 mm}
This paper proposes NeuroMAX, a high throughput accelerator using multi-threaded, log-based PE cores. The designed PE cores are capable of providing a 200\% increase in peak throughput while only increasing the area overhead by $6\%$, when compared to a standard multiplier-based PE core. We also design an efficient 2D weight broadcast dataflow scheme which exploits the multi-level parallelism of our processing engine and enables hardware utilization close to $100\%$. The accelerator is capable of performing a wide variety of convolutions including standard and separable $3\times3$ stride 1 and 2, $4\times4$, $5\times5$ and $1\times1$ depthwise, required in modern CNN architectures. We have implemented NeuroMAX on Xilinx Zynq-7020 SoC and have evaluated various performance parameters. The design can provide at least a throughput increase of $77.4\%$ and a latency decrease of $47\%$ with a $28\%$ decrease in PE count against recently proposed accelerator designs for modern CNNs. NeuroMAX also provides at least a $27\%$ and a $29\%$ decrease in power consumption and LUT count, respectively, against prior FPGA-based CNN accelerators.   

\balance
\bibliographystyle{unsrt}
\bibliography{bibv1.bib}

\end{document}